\input harvmac.tex

\def\sssec#1{$\underline{\rm #1}$}


\def\unlockat{\catcode`\@=11}
\def\lockat{\catcode`\@=12}

\unlockat

\def\newsec#1{\global\advance\secno  
by1\message{(\the\secno. #1)}
\global\subsecno=0\global\subsubsecno=0\eqnres@t\noindent
{\bf\the\secno. #1}
\writetoca{{\secsym} {#1}}\par\nobreak\medskip\nobreak}
\global\newcount\subsecno \global\subsecno=0
\def\subsec#1{\global\advance\subsecno
by1\message{(\secsym\the\subsecno. #1)}
\ifnum\lastpenalty>9000\else\bigbreak\fi\global\subsubsecno=0
\noindent{\it\secsym\the\subsecno. #1}
\writetoca{\string\quad {\secsym\the\subsecno.} {#1}}
\par\nobreak\medskip\nobreak}
\global\newcount\subsubsecno \global\subsubsecno=0
\def\subsubsec#1{\global\advance\subsubsecno by1
\message{(\secsym\the\subsecno.\the\subsubsecno. #1)}
\ifnum\lastpenalty>9000\else\bigbreak\fi
\noindent\quad{\secsym\the\subsecno.\the\subsubsecno.}{#1}
\writetoca{\string\qquad{\secsym\the\subsecno.\the\subsubsecno.}{#1}}
\par\nobreak\medskip\nobreak}

\def\subsubseclab#1{\DefWarn#1\xdef
#1{\noexpand\hyperref{}{subsubsection}%
{\secsym\the\subsecno.\the\subsubsecno}%
{\secsym\the\subsecno.\the\subsubsecno}}%
\writedef{#1\leftbracket#1}\wrlabeL{#1=#1}}
\lockat

\def\IB{\relax\hbox{$\inbar\kern-.3em{\rm B}$}}
\def\IC{\relax\hbox{$\inbar\kern-.3em{\rm C}$}}
\def\ID{\relax\hbox{$\inbar\kern-.3em{\rm D}$}}
\def\IE{\relax\hbox{$\inbar\kern-.3em{\rm E}$}}
\def\IF{\relax\hbox{$\inbar\kern-.3em{\rm F}$}}
\def\IG{\relax\hbox{$\inbar\kern-.3em{\rm G}$}}
\def\IGa{\relax\hbox{${\rm I}\kern-.18em\Gamma$}}
\def\IH{\relax{\rm I\kern-.18em H}}
\def\IK{\relax{\rm I\kern-.18em K}}
\def\IL{\relax{\rm I\kern-.18em L}}
\def\IP{\relax{\rm I\kern-.18em P}}
\def\IR{\relax{\rm I\kern-.18em R}}
\def\IZ{\relax\ifmmode\mathchoice
{\hbox{\cmss Z\kern-.4em Z}}{\hbox{\cmss Z\kern-.4em Z}}
{\lower.9pt\hbox{\cmsss Z\kern-.4em Z}}
{\lower1.2pt\hbox{\cmsss Z\kern-.4em Z}}\else{\cmss  
Z\kern-.4em
Z}\fi}
\def\CA{{\cal A}}

\def\CE {{\cal E}}
\def\CF {{\cal F}}

\def\CL {{\cal L}}
\def\CM {{\cal M}}

\def\CO {{\cal O}}
\def\CP {{\cal P}}

\def\CR {{\cal R}}


\def\p{\partial}



\def\Tr{\rm Tr}

\font\manual=manfnt  
\def\dbend{\lower3.5pt\hbox{\manual\char127}}

\def\c{\cdot}
\def\half {{1\over 2}}
\def\ch{{\rm ch}}

\def\inbar{\,\vrule height1.5ex width.4pt depth0pt}

\def\lieg{{\underline{\bf g}}}

\def\liet{{\underline{\bf t}}}



\font\cmss=cmss10 \font\cmsss=cmss10 at 7pt

\def\boxit#1{\vbox{\hrule\hbox{\vrule\kern8pt
\vbox{\hbox{\kern8pt}\hbox{\vbox{#1}}\hbox{\kern8pt}}
\kern8pt\vrule}\hrule}}
\def\mathboxit#1{\vbox{\hrule\hbox{\vrule\kern8pt\vbox{\kern8pt
\hbox{$\displaystyle #1$}\kern8pt}\kern8pt\vrule}\hrule}}


\def\inbar{\,\vrule height1.5ex width.4pt depth0pt}

\font\cmss=cmss10 \font\cmsss=cmss10 at 7pt


\def\a1{{\cal A}^{1,1}}

\def\hi{\chi^{2,0}}

%

\lref\ginzburg{V. Ginzburg, M. Kapranov, and E. Vasserot,
``Langlands Dualtiy for Surfaces,'' IAS preprint}

\lref\gottsh{L. Gottsche, Math. Ann. 286 (1990)193}
\lref\gothuy{L. G\"ottsche and D. Huybrechts,
``Hodge numbers of moduli spaces of stable
bundles on $K3$ surfaces,'' alg-geom/9408001}
\lref\GrHa{P.~ Griffiths and J.~ Harris, {\it Principles  
of Algebraic
geometry},
p. 445, J.Wiley and Sons, 1978. }
\lref\ripoff{I. Grojnowski, ``Instantons and
affine algebras I: the Hilbert scheme and
vertex operators,'' alg-geom/9506020.}
\lref\adhmfk{I. Grojnowski,
A. Losev, G. Moore, N. Nekrasov, S. Shatashvili,
``ADHM and the Frenkel-Kac construction,'' in preparation}

\lref\hirz{F. Hirzebruch and T. Hofer, Math. Ann. 286  
(1990)255}

\lref\milnor{J. Milnor, ``A unique decomposition
theorem for 3-manifolds,'' Amer. Jour. Math, (1961) 1}
\lref\ffeta{S.~ Ferrara et al., ``Prepotential and  
Monodromies in
$N=2$
Heterotic String,'' hep-th/9504034.}
\lref\fhsv{S.~ Ferrara,  J.A.~ Harvey,  A.~ Strominger,   
C.~ Vafa,
``Second-Quantized Mirror Symmetry,''
hep-th/9505162.}

\lref\afgnti{
I. Antoniadis, S. Ferrara, E. Gava, K.S. Narain, and
T.R. Taylor,
``Duality Symmetries in $N=2$ Heterotic Superstring,''
hep-th/9510079}
\lref\AFT{I. Antoniadis, S.Ferrara, T.R.Taylor, ``
$N=2$ Heterotic Superstring and its Dual Five Dimensional
Theory'', hep-th/9511108}
\lref\aspinwall{P.S.~ Aspinwall,  J.~ Louis,
``On the Ubiquity of K3 Fibrations in String Duality,''
hep-th/9510234}
\lref\vrldy{R.~ Dijkgraaf, E.~ Verlinde and H.~ Verlinde,
``Counting Dyons in $N=4$ String Theory'', CERN-TH/96-170,
hepth/9607}
\lref\vrlsq{E. Verlinde and H. Verlinde,
``Conformal Field Theory and Geometric Quantization,''
in {\it Strings '89},Proceedings
of the Trieste Spring School on Superstrings,
3-14 April 1989, M. Green, et. al. Eds. World
Scientific, 1990}
\lref\vrldglo{E. Verlinde, ``Global Aspects of
Electric-Magnetic Duality,'' hep-th/9506011}
\lref\ver{E. Verlinde, Nucl. Phys. {\bf B} 300 (1988) 360}
\lref\dvvm{R. Dijkgraaf, G, Moore, E. Verlinde, H. Verlinde,
``Elliptic Genera of Symmetric Products and Second Quantized
Strings'',
hep-th/9608096, CERN-TH/96-222, ITFA/96-31, YCTP-P16-96}

\lref\gerasimov{A.~ Gerasimov, ``Localization in GWZW and  
Verlinde
formula'', UUITP 16/1993, hep-th/9305090}
\lref\BlThlgt{M.~ Blau and G.~ Thompson, ``Lectures on 2d  
Gauge
Theories: Topological Aspects and Path
Integral Techniques", Presented at the
Summer School in Hogh Energy Physics and
Cosmology, Trieste, Italy, 14 Jun - 30 Jul
1993, hep-th/9310144.}
\lref\btverlinde{M.~ Blau, G.~ Thomson,
``Derivation of the Verlinde Formula from Chern-Simons  
Theory and the
$G/G$
   model'',Nucl. Phys. {\bf B}408 (1993) 345-390 }
\lref\btqmech{M.~ Blau , G.~ Thompson, ``Topological  
gauge theories
from supersymmetric quantum mechanics on
spaces of connections'', Int. J. Mod. Phys. A8 (1993) 573-586}
\lref\Candelas{P. Candelas, X. De la Ossa, A. Font, S.  
Katz, and D.
Morrison,
``Mirror Symmetry for Two Parameter Models - I,'' {\it  
Nucl. Phys.}
{\bf B416} (1994) 481, hep-th/9308083.}

\lref\Cardoso{ Gabriel Lopes Cardoso,  Gottfried Curio,   
Dieter Lust,
Thomas Mohaupt,  Soo-Jong Rey,
``BPS Spectra and Non--Perturbative Couplings in N=2,4  
Supersymmetric
String
Theories,''
hep-th/9512129}

\lref\gn{Valeri A. Gritsenko  ,  Viacheslav V. Nikulin,
``
K3 surfaces, Lorentzian Kac--Moody algebras and Mirror  
Symmetry,''
alg-geom/9510008}
\lref\diss{N.~ Nekrasov, PhD. Thesis, Princeton 1996}
\lref\galich{N.~ Nekrasov,
``Higher Verlinde Formulas'', to appear}
\lref\hm{ J. A. Harvey  ,  G. Moore,
``Algebras, BPS States, and Strings,''
hep-th/9510182}
\lref\bost{L. Alvarez-Gaume, J.B. Bost , G. Moore, P.  
Nelson, C.
Vafa,
``Bosonization on higher genus Riemann surfaces,''
Commun.Math.Phys.112:503,1987}
\lref\agmv{L. Alvarez-Gaum\'e,
C. Gomez, G. Moore,
and C. Vafa, ``Strings in the Operator Formalism,''
Nucl. Phys. {\bf 303}(1988)455}
\lref\hms{hep-th/9501022,
 Reducing $S$- duality to $T$- duality, J. A. Harvey, G.  
Moore and A.
Strominger}

\lref\lomns{A. Losev, G. Moore, N. Nekrasov,
S. Shatashvili, unpublished.}.
\lref\CMR{ For a review, see
S. Cordes, G. Moore, and S. Ramgoolam,
`` Lectures on 2D Yang Mills theory, Equivariant
Cohomology, and Topological String Theory,''
Lectures presented at the 1994 Les Houches Summer School
 ``Fluctuating Geometries in Statistical Mechanics and Field
Theory.''
and at the Trieste 1994 Spring school on superstrings.
hep-th/9411210, or see http://xxx.lanl.gov/lh94}
\lref\elitzur{S. Elitzur, G. Moore,
A. Schwimmer, and N. Seiberg,
``Remarks on the Canonical Quantization of the Chern-Simons-
Witten Theory,'' Nucl. Phys. {\bf B326}(1989)108 \semi
G. Moore and N. Seiberg,
``Lectures on Rational Conformal Field Theory'',
in {\it Strings'89},Proceedings
of the Trieste Spring School on Superstrings,
3-14 April 1989, M. Green, et. al. Eds. World
Scientific, 1990}
\lref\adhmfk{I. Grojnowski,
A. Losev, G. Moore, N. Nekrasov, S. Shatashvili,
``ADHM and the Frenkel-Kac construction,'' in preparation}
\lref\hypvol{A. Losev, G. Moore, N. Nekrasov, S. Shatashvili,
``Localization for Hyperkahler Quotients,
Integration over Instanton Moduli,
and ALE Spaces,'' in preparation}
\lref\fdrcft{A. Losev, G. Moore, N. Nekrasov, S.  
Shatashvili, in
preparation.}
\lref\cenexts{A. Losev, G. Moore, N. Nekrasov, S. Shatashvili,
``Central Extensions of Gauge Groups Revisited,''
hep-th/9511185.}
\lref\taming{G. Moore and N. Seiberg,
``Taming the conformal zoo,'' Phys. Lett.
{\bf 220 B} (1989) 422}

\lref\KLM{A. Klemm, W. Lerche, and P. Mayr,  
``K3-Fibrations and
Heterotic-Type II String Duality,'' hep-th/9506112.}

\lref\KV{S. Kachru and C. Vafa,
`` Exact Results for N=2 Compactifications of Heterotic  
Strings,''
hep-th/9505105; Nucl. Phys. B450 (1995) 69-89}
\lref\KKLMV{
S. Kachru, A. Klemm, W. Lerche, P. Mayr and C. Vafa,
``Nonperturbative Results on the Point Particle
 Limit of N=2 Heterotic String Compactifications,''
hep-th/9508155.
}
\lref\kawai{Toshiya Kawai,
``$N=2$ heterotic string threshold correction, $K3$
surface and generalized Kac-Moody superalgebra,''
hep-th/9512046
}
\lref\HET{D. Gross, J. Harvey, E. Martinec, and R. Rohm, ``The
Heterotic String,'' {\it Phys. Rev. Lett.} {\bf {54}}  
(1985) 502.}
\lref\grjakiw{D.J.~ Gross, R.~ Jackiw, ....
``Towards the Theory of Strong Interactions''}
\lref\grwilczek{D.J.~ Gross, F.~ Wilczek,  
``Asymptotically Free
Gauge Theories. 1'', Phys. Rev. D8 (1973) 3633-3652, \semi
``Asymptotically Free
Gauge Theories. 2'', Phys.Rev.D9:980-993,1974. }
\lref\grcurrent{D.J.~ Gross, C.H.~ Lewellyn Smith,  
``High-Energy
Neutrino-Nucleon Scattering, Current Algebra and Partons'',
Nucl.Phys.B14:337-347,1969.}
\lref\grcoleman{
S.~ Coleman,
R.~ Jackiw, D.J.~ Gross, ``Fermion
Avatars of Sugawara Model''}
\lref\recqcd{W.~ A.~ Bardeen, ``Self-Dual Yang-Mills,  
Integrability
and Multi-Parton Amplitudes'',
Fermilab - Conf - -95-379-T, Aug 1995,
Presented at Yukawa International Seminar '95:
`From the Standard Model to Grand
Unified Theories', Kyoto, Japan, 21-25 Aug 1995. \semi
D. Cangemi, ``Self-dual Yang-Mills
Theory and One-Loop Like-Helicity QCD
Multi-Gluon Amplitudes,'' hep-th/9605208.}
\lref\gwdzki{K. Gawedzki, ``Topological Actions in  
Two-Dimensional
Quantum Field Theories,'' in {\it Nonperturbative
Quantum Field Theory}, G. 't Hooft, A. Jaffe, et. al. , eds. ,
Plenum 1988}

\lref\shatashi{S. Shatashvili,
Theor. and Math. Physics, 71, 1987, p. 366}
\lref\gmps{A. Gerasimov, A. Morozov, M. Olshanetskii,
 A. Marshakov, S. Shatashvili, ``Wess-Zumino-Witten model  
as a theory
of
free fields,'' Int. J. Mod. Phys. A5 (1990) 2495-2589}
\lref\frolov{G.E.~ Arutyunov, S.A.~ Frolov and P.B.~ Medvedev,
``Elliptic Ruijsenaars-Schneider model from the cotangent  
bundle over
the
  two-dimensional current group'',  hep-th/9608013}
\lref\vafagas{C. Vafa,
``Gas of D-Branes and Hagedorn Density of BPS States'',
hep-th/9511088}
\lref\vafa{C. Vafa, ``Conformal theories and punctured
surfaces,'' Phys.Lett.199B:195,1987 }
\lref\VaWi{C.~ Vafa and E.~ Witten, ``A Strong Coupling  
Test of
$S$-Duality",
hep-th/9408074.}
\lref\giveon{hep-th/9502057,
 S-Duality in N=4 Yang-Mills Theories with General Gauge  
Groups,
 Luciano Girardello, Amit Giveon, Massimo Porrati, and Alberto
Zaffaroni
}

\lref\OldLG{C. Vafa, ``String Vacua and Orbifoldized LG  
Models,''
{\it Mod. Phys. Lett.} {\bf A4} (1989) 1169 \semi
K. Intriligator and C. Vafa, ``Landau-Ginzburg Orbifolds,"
{\it Nucl. Phys.}{\bf B339} (1990) 95 \semi
S. Cecotti, L. Girardello, and A. Pasquinucci,
``Non-perturbative Aspects and Exact Results for the $N=2$
Landau-Ginzburg
Models,'' Nucl. Phys. B338 (1989) 701, ``Singularity
Theory and N=2 Supersymmetry,'' Int. J. Mod. Phys. {\bf  
A6} (1991)
2427.}
\lref\bcov{M.~ Bershadsky, S.~ Cecotti, H.~ Ooguri and  
C.~ Vafa,
``Kodaira-Spencer Theory of Gravity and Exact Results for  
Quantum
String Amplitudes'', Comm. Math. Phys. 165(1994):311-428}
\lref\bjsv{M.~ Bershadsky, A.~ Johansen, V.~ Sadov and  
C.~ Vafa,
``Topological Reduction of 4D SYM to 2D $\sigma$--Models'',
hep-th/9501096,}
\lref\ogvf{H. Ooguri and C. Vafa, ``Self-Duality
and $N=2$ String Magic,'' Mod.Phys.Lett. {\bf A5} (1990)
1389-1398\semi
``Geometry
of$N=2$ Strings,'' Nucl.Phys. {\bf B361}  (1991) 469-518.}
\lref\VafaQ{C. Vafa, ``Quantum Symmetries of String Vacua,''
Mod. Phys. Lett. {\bf A4} (1989) 1615. }
\lref\Vafa{C. Vafa, ``String Vacua and Orbifoldized LG  
models,''
{\it Mod. Phys. Lett.} {\bf A4} (1989) 1169.}
\lref\Ken{K. Intriligator and C. Vafa, ``Landau-Ginzburg  
Orbifolds,''
{\it Nucl. Phys.} {\bf B339} (1990) 95.}
\lref\berk{N. Berkovits,
``Super-Poincare Invariant Superstring Field Theory''
hep-th/9503099 }

\lref\SING{C. Vafa, ``Strings and Singularities,'' Harvard
preprint, {\rm hep-th/9310069}.}
\lref\dkv{M.~ Douglas, S.~ Katz and C.~ Vafa, ``Small  
Instantons, Del
Pezzo
Surfaces and Type $I^{\prime}$ Theory'', hep-th/9609071,
HUTP-96/A042,
RU-96-79, OSU-M-96-22}
\lref\WitDonagi{
R.~ Donagi, E.~ Witten, ``Supersymmetric Yang-Mills Theory and
Integrable Systems'', hep-th/9510101, Nucl.Phys.{\bf
B}460:299-334,1996}
\lref\Witfeb{E.~ Witten, ``Supersymmetric Yang-Mills  
Theory On A
Four-Manifold,'' J. Math. Phys. {\bf 35} (1994) 5101.}
\lref\Witr{E.~ Witten, ``Introduction to Cohomological Field
Theories",
Lectures at Workshop on Topological Methods in Physics,  
Trieste,
Italy,
Jun 11-25, 1990, Int. J. Mod. Phys. {\bf A6} (1991) 2775.}
\lref\Witgrav{E.~ Witten, ``Topological Gravity'',
Phys.Lett.206B:601,1988}
\lref\witaffl{I. ~ Affleck, J.A.~ Harvey and E.~ Witten,
	``Instantons and (Super)Symmetry Breaking
	in $2+1$ Dimensions'', Nucl. Phys. {\bf B206}  
(1982) 413}
\lref\witlec{E.~ Witten, lectures at IAS, fall 1995}
\lref\wittabl{E.~ Witten,  ``On S-Duality in Abelian  
Gauge Theory,''
hep-th/9505186}
\lref\wittgr{E.~ Witten, ``The Verlinde Algebra And The  
Cohomology Of
The Grassmannian'',  hep-th/9312104}
\lref\wittenwzw{E. Witten, ``Nonabelian bosonization in
two dimensions,'' Commun. Math. Phys. {\bf 92} (1984)455 }
\lref\witgrsm{E. Witten, ``Quantum field theory,
grassmannians and algebraic curves,''  
Commun.Math.Phys.113:529,1988}
\lref\wittjones{E. Witten, ``Quantum field theory and the  
Jones
polynomial,'' Commun.  Math. Phys., 121 (1989) 351. }
\lref\witttft{E.~ Witten, ``Topological Quantum Field Theory",
Commun. Math. Phys. {\bf 117} (1988) 353.}
\lref\wittmon{E.~ Witten, ``Monopoles and Four-Manifolds'',
hep-th/9411102}
\lref\Witdgt{ E.~ Witten, ``On Quantum gauge theories in two
dimensions,''
Commun. Math. Phys. {\bf  141}  (1991) 153\semi
 ``Two dimensional gauge
theories revisited'', J. Geom. Phys. 9 (1992) 303-368}
\lref\Witgenus{E.~ Witten, ``Elliptic Genera and Quantum Field
Theory'',
Comm. Math. Phys. 109:525,1987. }

\lref\OldWit{E. Witten, ``New Issues in Manifolds of SU(3)
Holonomy,''
 {\it Nucl. Phys.} {\bf B268} (1986) 79.}
\lref\OldZT{E. Witten, ``New Issues in Manifolds of SU(3)  
Holonomy,''
{\it Nucl. Phys.} {\bf B268} (1986) 79 \semi
J. Distler and B. Greene, ``Aspects of (2,0) String
Compactifications,''
{\it Nucl. Phys.} {\bf B304} (1988) 1 \semi
B. Greene, ``Superconformal Compactifications in Weighted  
Projective
Space,'' {\it Comm. Math. Phys.} {\bf 130} (1990) 335.}
\lref\bagger{E.~ Witten and J. Bagger, Phys. Lett.
{\bf 115B}(1982) 202}
\lref\witcurrent{E.~ Witten,``Global Aspects of Current  
Algebra'',
Nucl.Phys.B223 (1983) 422\semi
``Current Algebra, Baryons and Quark Confinement'',  
Nucl.Phys. B223
(1993)
433}
\lref\Wittreiman{S.B. Treiman,
E. Witten, R. Jackiw, B. Zumino, ``Current Algebra and
Anomalies'', Singapore, Singapore: World Scientific ( 1985) }
\lref\Witgravanom{L. Alvarez-Gaume, E.~ Witten,  
``Gravitational
Anomalies'',
Nucl.Phys.B234:269,1984. }

\lref\CHSW{P. Candelas, G. Horowitz, A. Strominger and E.  
Witten,
``Vacuum Configurations for Superstrings,'' {\it Nucl.  
Phys.} {\bf
B258} (1985) 46.}
\lref\AandB{E. Witten, in ``Proceedings of the Conference  
on Mirror
Symmetry",
MSRI (1991).}
\lref\Witr{E.~ Witten, ``Introduction to Cohomological Field
Theories",
Lectures at Workshop on Topological Methods in Physics,  
Trieste,
Italy,
Jun 11-25, 1990, Int. J. Mod. Phys. {\bf A6} (1991) 2775.}

\lref\phases{E. Witten, ``Phases of N=2 Theories in Two  
Dimensions",
{\it Nucl. Phys.} {\bf B403} (1993) 159, {\rm  
hep-th/9301042}.}
\lref\Us{S. Kachru and E. Witten, ``Computing The  
Complete Massless
Spectrum Of A Landau-Ginzburg Orbifold,''
{\it Nucl. Phys.} {\bf B407} (1993) 637, {\rm  
hep-th/9307038}.}
\lref\WitMin{E. Witten,
``On the Landau-Ginzburg Description of N=2 Minimal Models,''
IAS preprint
IASSNS-HEP-93/10, hep-th/9304026.}
\lref\witremarks{E.~ Witten, ``Some Comments on String  
Dynamics'',
hep-th/9507121}
\lref\wittrans{E.~ Witten, ``Phase Transitions in $M$-Theory
and $F$-Theory'', hep-th/9603150}
\lref\twisted{E. Witten. Comm. Math. Phys. {\bf 118}  
(1988) 411\semi
E. Witten, Nucl. Phys. {\bf B340} (1990) 281\semi
T. Eguchi and S.-K. Yang,
Mod. Phys. Lett. {\bf A5} (1990) 1693.}
\lref\witseiii{N.~ Seiberg, E.~ Witten,
``Gauge Dynamics and Compactification to Three Dimensions,''
IASSNS-HEP-96-78, RU-96-55, hep-th/9607163}

\lref\KLT{ Vadim Kaplunovsky  ,  Jan Louis  ,  Stefan Theisen,
``Aspects of Duality in N=2 String Vacua,''
hep-th/9506110;
{\it Phys. Lett.} {\bf B357} (1995) 71.
}

\lref\dWKLL{B. de Wit, V. Kaplunovsky, J. Louis, and D. Lust,
``Perturbative Couplings of Vector Multiplets in N=2  
Heterotic String
Vacua,''
hep-th/9504006.}
\lref\Yau{S. Hosono, A. Klemm, S. Theisen, and S.T. Yau,  
``Mirror
Symmetry,
Mirror Map and Applications to Calabi-Yau Hypersurfaces,''
Comm. Math. Phys. {\bf 167} (1995) 301, hep-th/9308122;
S. Hosono, A. Klemm, S. Theisen, and S.T. Yau,
``Mirror Symmetry, Mirror Map, and Applications to Complete
Intersection
Calabi-Yau Spaces,''
{\it Nucl. Phys.} {\bf B433} (1995) 501, hep-th/9406055.}

\lref\HarStro{J.~ Harvey, A.~ Strominger, ``String Theory  
and the
Donaldson  Polytnomial'',
hep-th/9108020, Comm. Math. Phys. 151(1993), 221-232}
\lref\blzh{A. Belavin, V. Zakharov, ``Yang-Mills  
Equations as inverse
scattering
problem''Phys. Lett. B73, (1978) 53}
\lref\bpz{A.A. Belavin, A.M. Polyakov, A.B. Zamolodchikov,
``Infinite conformal symmetry in two-dimensional quantum
field theory,'' Nucl.Phys.B241:333,1984}

\lref\atiyah{M. Atiyah, ``Green's Functions for
Self-Dual Four-Manifolds,'' Adv. Math. Suppl.
{\bf 7A} (1981)129}
\lref\AHS{M.~ Atiyah, N.~ Hitchin and I.~ Singer,
``Self-Duality in Four-Dimensional
Riemannian Geometry", Proc. Royal Soc. (London) {\bf  
A362} (1978)
425-461.}
\lref\fmlies{M.~ Atiyah and I.~ Singer,
``The index of elliptic operators IV,'' Ann. Math. {\bf  
93}(1968)119}

\lref\hitchin{N.~ Hitchin, ``Polygons and gravitons,''
Math. Proc. Camb. Phil. Soc, (1979){\bf 85} 465}
\lref\hklr{Hitchin, Karlhede, Lindstrom, and Rocek,
``Hyperkahler metrics and supersymmetry,''
Commun. Math. Phys. {\bf 108}(1987)535}
\lref\hi{N.~ Hitchin, Duke Math. Journal, Vol. 54, No. 1  
(1987)}


\lref\banks{T. Banks, ``Vertex Operators in 2D Dimensions,''
hep-th/9503145}

\lref\donii{
S. Donaldson, ``Infinite Determinants, Stable
Bundles, and Curvature,''
Duke Math. J. , {\bf 54} (1987) 231. }
\lref\BGS{J. -M. Bismut, H. Gillet, and C. Soul\'e,
``Analytic Torsion and Holomorphic Determinant
Bundles, I.II.III''
CMP {\bf 115}(1988)49-78;79-126;301-351}
\lref\RS{D.~ Ray and I.M.~ Singer, ``R-torsion and the  
Laplasian on
Riemannian Manifolds'', Adv. in Math. 7 (1971), 145-210}
\lref\quillen{D.~ Quillen, ``Determinants of  
Cauchy-Riemann operators
over a Riemann Surface'', Funk. Anal. i Prilozen. 19,  
37-41 (1985) [=
Func. Anal. Appl. 19, 31-34 (1986)]}
\lref\bismut{J.-M.~ Bismut, ``The Atyah-Singer theorem  
for families
of Dirac Operators: two heat kernel proofs'', Invent.  
Math. 83,
91-151
(1986)}
\lref\BottChern{R.~ Bott and S.~ Chern, ``Hermitian  
Vector Bundles
and the Equidistribution of the zeroes of their holomorphic
cross sections'', Acta Math. 114, 71-112 (1968)}
\lref\braam{P.J. Braam, A. Maciocia, and A. Todorov,
``Instanton moduli as a novel map from tori to
K3-surfaces,'' Inven. Math. {\bf 108} (1992) 419}
\lref\maciocia{A. Maciocia, ``Metrics on the moduli
spaces of instantons over Euclidean 4-Space,''
Commun. Math. Phys. {\bf 135}(1991) , 467}
\lref\martinecsix{E.~ Martinec, N.~ Warner, ``Integrability
in $N=2$ Theories: A Proof'', hep-th/9511052}
\lref\cllnhrvy{Callan and Harvey, Nucl. Phys. {\bf  
B250}(1985)427}
\lref\galperin{A. Galperin, E. Ivanov, V. Ogievetsky,
E. Sokatchev, Ann. Phys. {\bf 185}(1988) 1}

\lref\evans{M. Evans, F. G\"ursey, V. Ogievetsky,
``From 2D conformal to 4D self-dual theories:
Quaternionic analyticity,''
Phys. Rev. {\bf D47}(1993)3496}

\lref\devchand{Ch. Devchand and V. Ogievetsky,
``Four dimensional integrable theories,'' hep-th/9410147}
\lref\devchandi{
Ch. Devchand and A.N. Leznov,
``B \"acklund transformation for supersymmetric self-dual  
theories
for
semisimple
gauge groups and a hierarchy of $A_1$ solutions,''  
hep-th/9301098,
Commun. Math. Phys. {\bf 160} (1994) 551}

\lref\evans{M. Evans, F. G\"ursey, V. Ogievetsky,
``From 2D conformal to 4D self-dual theories:
Quaternionic analyticity,''
Phys. Rev. {\bf D47}(1993)3496}
\lref\fs{L.~ Faddeev and S.~ Shatashvili, Theor. Math.  
Fiz., 60
(1984)
206}
\lref\faddeevlmp{L.D.~ Faddeev, ``Some Comments on Many  
Dimensional
Solitons'',
Lett. Math. Phys., 1 (1976) 289-293.}
\lref\nov{ S. Novikov,"The Hamiltonian
formalism and many-valued analogue of  Morse theory",
Russian Math.Surveys 37:5(1982),1-56}
\lref\fsi{L.~ Faddeev, Phys. Lett. B145 (1984) 81.}
\lref\mick{I. Mickellson, CMP, 97 (1985) 361.}

\lref\fz{I.~ Frenkel, I.~ Singer, unpublished.}
\lref\fk{I.~ Frenkel and B.~ Khesin, ``Four dimensional
realization of two dimensional current groups,'' Yale
preprint, July 1995, to appear in Commun. Math. Phys.}
\lref\ff{B. Feigin , E. Frenkel, N. Reshetikhin, {\it  
Gaudin Model,
Critical Level and Bethe Ansatz}, CMP {\bf  166} (1995),  
27-62}
\lref\efk{P.~ Etingof, I.~ Frenkel, A.~ Kirillov,Jr.,
``Spherical functions on affine Lie groups'', Yale  
preprint, 1994}
\lref\etingof{P.I. Etingof and I.B. Frenkel,
``Central Extensions of Current Groups in
Two Dimensions,'' Commun. Math.
Phys. {\bf 165}(1994) 429}
\lref\dnld{S. Donaldson, ``Anti self-dual Yang-Mills
connections over complex  algebraic surfaces and stable
vector bundles,'' Proc. Lond. Math. Soc,
{\bf 50} (1985)1}
\lref\biquard{O. Biquard, ``Sur les fibr\'es paraboliques
sur une surface complexe,'' to appear in J. Lond. Math.
Soc.}

\lref\DoKro{S.K.~ Donaldson and P.B.~ Kronheimer,
``The Geometry of Four-Manifolds'',
Clarendon Press, Oxford, 1990.}
\lref\donii{
S. Donaldson, Duke Math. J. , {\bf 54} (1987) 231. }


\lref\ShiBeta{V.~ Novikov, M.A. Shifman, A.I. Vainshtein, V.I.
Zakharov,
``Exact  Gell-Mann-Low Function of Supersymmetric Yang-Mills
Theories From Instanton
Calculus'', Nucl.Phys.B229:381,1983. \semi
``Beta Function in Supersymmetric Gauge Theories:
Instantons Versus Traditional Approach'',  
Phys.Lett.166B:329,1986}
\lref\gwdzki{K.~ Gawedzki, ``Topological Actions in  
Two-Dimensional
Quantum Field Theories,'' in {\it Nonperturbative
Quantum Field Theory}, G. 't Hooft, A. Jaffe, et. al. , eds. ,
Plenum 1988}
\lref\ginzburg{V. Ginzburg, M. Kapranov, and E. Vasserot,
``Langlands Reciprocity for Algebraic Surfaces,''  
q-alg/9502013}
\lref\giveon{
 ``S-Duality in N=4 Yang-Mills Theories with General  
Gauge Groups'',
hep-th/9502057,
 Luciano Girardello, Amit Giveon, Massimo Porrati, and Alberto
Zaffaroni
}
\lref\mickelsson{J. Mickelsson, ``Kac-Moody groups,  
topology of
the Dirac  determinant bundle and fermionization,''
Commun. Math. Phys. {\bf 110}(1987)173.}
\lref\gottsh{L. Gottsche, Math. Ann. 286 (1990)193}
\lref\gothuy{L. G\"ottsche and D. Huybrechts,
``Hodge numbers of moduli spaces of stable
bundles on $K3$ surfaces,'' alg-geom/9408001}
\lref\GrHa{P.~ Griffiths and J.~ Harris, ``Principles of
Algebraic
geometry'',
p. 445, J.Wiley and Sons, 1978. }
\lref\ripoff{I. Grojnowski, ``Instantons and
affine algebras I: the Hilbert scheme and
vertex operators,'' alg-geom/9506020.}
\lref\hirz{F. Hirzebruch and T. Hofer, Math. Ann. 286  
(1990)255}

\lref\johansen{A. Johansen, ``Infinite Conformal
Algebras in Supersymmetric Theories on
Four Manifolds,'' hep-th/9407109, Nucl. Phys. B436 (1995)  
291-341
\semi
``Realization of $W_{1+\infty}$ and Virasoro Algebras in
Supersymmetric
   Theories on
Four Manifolds'',
hep-th/9406156, Mod. Phys. Lett. A9 (1994) 2611-2622\semi
``Twisting of
$N=1$ SUSY Gauge
Theories and Heterotic Topological Theories'',hep-th/9403017}
\lref\kronheimer{P. Kronheimer, ``The construction of ALE  
spaces as
hyper-kahler quotients,'' J. Diff. Geom. {\bf 28}1989)665}
\lref\kricm{P. Kronheimer, ``Embedded surfaces in
4-manifolds,'' Proc. Int. Cong. of
Math. (Kyoto 1990) ed. I. Satake, Tokyo, 1991}
\lref\krmw{P. Kronheimer and T. Mrowka,
``Gauge theories for embedded surfaces I,''
Topology {\bf 32} (1993) 773,
``Gauge theories for embedded surfaces II,''
preprint.}
\lref\rade{J. Rade, ``Singular Yang-Mills fields. Local
theory I. '' J. reine ang. Math. , {\bf 452}(1994)111;  
{\it ibid}
{\bf 456}(1994)197; ``Singular Yang-Mills
fields-global theory,'' Intl. J. of Math. {\bf 5}(1994)491.}
\lref\biquard{O. Biquard, ``Sur les fibr\'es paraboliques
sur une surface complexe,'' to appear in J. Lond. Math.
Soc.}
\lref\KN{P.~ Kronheimer and H.~ Nakajima,  ``Yang-Mills  
instantons
on ALE gravitational instantons,''  Math. Ann.
{\bf 288}(1990)263}
\lref\nakajima{H. Nakajima, ``Homology of moduli
spaces of instantons on ALE Spaces. I'' J. Diff. Geom.
{\bf 40}(1990) 105; ``Instantons on ALE spaces,
quiver varieties, and Kac-Moody algebras,'' preprint,
``Gauge theory on resolutions of simple singularities
and affine Lie algebras,'' preprint.}
\lref\nakheis{H.Nakajima, ``Heisenberg algebra and  
Hilbert schemes of
points on
projective surfaces ,'' alg-geom/9507012}

\lref\mickold{J. Mickelsson, CMP, 97 (1985) 361.}
\lref\milnor{J. Milnor, ``A unique decomposition
theorem for 3-manifolds,'' Amer. Jour. Math, (1961) 1}
\lref\milnsta{J. Milnor, J.Stasheff, ``Characteristic  
Classes'',
Princeton University Press, Princeton, New Jersey, USA, 1974}

\lref\nair{V.P.Nair, ``K\"ahler-Chern-Simons Theory'',
hep-th/9110042}
\lref\ns{V.P. Nair and Jeremy Schiff,
``Kahler Chern Simons theory and symmetries of
antiselfdual equations'' Nucl.Phys.B371:329-352,1992;
``A Kahler Chern-Simons theory and quantization of the
moduli of antiselfdual instantons,''
Phys.Lett.B246:423-429,1990,
``Topological gauge theory and twistors,''
Phys.Lett.B233:343,1989}

\lref\park{J.-S. Park, ``Holomorphic Yang-Mills theory on  
compact
Kahler
manifolds,'' hep-th/9305095; Nucl. Phys. {\bf B423}  
(1994) 559;
J.-S.~ Park, ``$N=2$ Topological Yang-Mills Theory on Compact
K\"ahler
Surfaces", Commun. Math, Phys. {\bf 163} (1994) 113;
S. Hyun and J.-S.~ Park, ``$N=2$ Topological Yang-Mills  
Theories
and Donaldson
Polynomials", hep-th/9404009}
\lref\parki{S. Hyun and J.-S. Park,
``Holomorphic Yang-Mills Theory and Variation
of the Donaldson Invariants,'' hep-th/9503036}
\lref\pohl{Pohlmeyer, Comm.
Math. Phys. {\bf 72}(1980)37}
\lref\prseg{A.~Pressley and G.~Segal, "Loop Groups",  
Oxford Clarendon
Press, 1986}
\lref\segal{G. Segal, The definition of CFT}
\lref\miracles{M. Dine and N. Seiberg, ``Are (0,2) Models
String Miracles?,"
{\it Nucl. Phys.} {\bf B306} (1988) 137.}
\lref\GrSei{M. Green and N. Seiberg, ``Contact interactions in
superstring theory," Nucl. Phys. {\bf B299} (1988) 559.}
\lref\DSWW{M. Dine, N. Seiberg, X.G. Wen and E. Witten,
``Non-Perturbative Effects on the String World Sheet I,''  
{\it Nucl.
Phys.}~{\bf B278} (1986) 769, ``Non-Perturbative Effects  
on the
String
World Sheet II,'' {\it Nucl. Phys.}~{\bf B289} (1987) 319. }
\lref\SeObserv{N.~  Seiberg, Nucl. Phys. {\bf B303}  
(1988) 286.}
\lref\SeBeta{N.~ Seiberg, ``Supersymmetry and
Non-perturbative Beta-Functions'', Phys.Lett.206B:75,1988}
\lref\SeWi{N. Seiberg, E. Witten, ``Electric-Magnetic Duality,
Monopole Condensation, And Confinement in $N=2$ Supersymmetric
Yang-Mills Theory ''
Nucl. Phys. B426 (1994) 19-52 (and erratum - ibid. B430 (1994)
485-486 )\semi
``Monopoles, Duality and Chiral Symmetry Breaking in
N=2 Supersymmetric QCD'', hep-th/9408099,
Nucl. Phys. B431 (1994) 484-550.}
\lref\seiberg{
	``Monopole Condensation and Confinement'',
hep-th/9408013,  Nathan Seiberg; hep-th/9408155,
Phases of N=1 supersymmetric gauge theories in four  
dimensions,
K.~ Intriligator
and N.~ Seiberg; hep-ph/9410203,
Proposal for a Simple Model of Dynamical SUSY Breaking,
by K.~ Intriligator, N.~ Seiberg, and S.~ H.~ Shenker;
hep-th/9411149,
 Electric-Magnetic Duality in Supersymmetric Non-Abelian Gauge
Theories,
 N. Seiberg; hep-th/9503179 Duality, Monopoles, Dyons,  
Confinement
and Oblique
Confinement in Supersymmetric $SO(N_c)$ Gauge Theories,
K. Intriligator and N. Seiberg}
\lref\morrseib{D.~ Morrison and N.~ Seiberg, ``Extremal  
Transitions
and Five-Dimensional Supersymmetric Gauge Theories'',  
hep-th/9609070,
DUKE-TH-96-130, RU-96-80}
\lref\seifive{N.~ Seiberg, ``Five Dimensional SUSY Field  
Theories,
Non-Trivial Fixed Points
and String Dynamics'', hep-th/ 9608111}
\lref\seibbr{N.~ Seiberg, ``Infrared Dynamics on Branes and
Space-Time Geometry'', hepth/9606017}
\lref\ganor{O.~ Ganor, ``Toroidal Compactification of  
Heterotic $6D$
Non-Critical Strings
Down to Four Dimensions'', hep-th/9608109}
\lref\argfar{P.C.~ Argyres, A.E.~ Farragi,
	``The Vacuum Structure and Spectrum
	of $N=2$ Supersymmetric $SU(n)$ Gauge Theory'',  
hep-th/9411057\semi
	A.~ Klemm, W.~ Lerche, S.~ Theisen and S.~  
Yankielowicz,
	``Simple Singularities and
	$N=2$ Supersymmetric Yang-Mills Theory'',  
hep-th/9411048}
\lref\argsei{P.C.~ Argyres, M.R.~ Plesser and N.~ Seiberg,
	``The Moduli Space of Vacua of
	$N=2$ SUSY QCD and Duality in $N=1$
	SUSY QCD'', hep-th/9603042}

\lref\sen{A. Sen,
 ``Dyon-Monopole bound states, selfdual harmonic
forms on the multimonopole moduli space and $SL(2,Z)$
invariance'', hep-th/9402032
}
\lref\thooft{G. 't Hooft , ``A property of electric and
magnetic flux in nonabelian gauge theories,''
Nucl.Phys.B153:141,1979}
\lref\wrdhd{R. Ward, Nucl. Phys. {\bf B236}(1984)381}
\lref\ward{Ward and Wells, {\it Twistor Geometry and
Field Theory}, CUP }

\lref\avatar{A.~ Losev, G.~ Moore, N.~ Nekrasov, S.~  
Shatashvili,
``Four-Dimensional Avatars of 2D RCFT,''
hep-th/9509151, Talks at Strings'95 and Trieste 1995}
\lref\cocycle{A.~ Losev, G.~ Moore, N.~ Nekrasov, S.~  
Shatashvili,
``Central Extensions of Gauge Groups Revisited,''
hep-th/9511185.}
\lref\clash{A.~ Losev, G.~ Moore, N.~ Nekrasov, S.~  
Shatashvili,
`` Chiral Lagrangians, Anomalies, Supersymmetry,
and Holomorphy'', PUPT-1627,ITEP-TH-18/96,YCTP-P10-96,
hep-th/9606082}


\lref\gnell{A. Gorsky, N. Nekrasov, ``Elliptic  
Calogero-Moser System
from
Two Dimensional Current Algebra'', hepth/9401021}
\lref\gnru{A.Gorsky, N.Nekrasov, ``Relativistic  
Calogero-Moser model
as
gauged WZW theory'', Nucl.Phys. {\bf B} 436 (1995) 582-608,
hep-th/9401017}
\lref\gnfu{N. Nekrasov, A. Gorsky ,
``Integrable systems on moduli
spaces'', in preparation}
\lref\dual{V.~ Fock, A.~ Gorsky, N.~ Nekrasov, V.~ Rubtsov,
``Duality in Many-Body Systems and Gauge Theories'', in  
preparation}

\lref\krzabr{I.~ Krichever, A.~ Zabrodin, ``Spin  
Generalization of
Ruijsenaars-Schneider model, Non-abelian Toda Chain and
Representations
of Sklyanin Algebra'', hep-th/9505039}
\lref\gmmm{
A.~ Gorsky,~ I.~ Krichever,~ A.~ Marshakov,~ A.~ Morozov,~ A.~
Mironov,~
``Integrablity and
Seiberg-Witten Exact Solution'',
hep-th/9505035,  Phys. Lett. B355: 466-474, 1995}
\lref\bbkt{O.~ Babelon, E.~ Billey, I.~ Krichever, M.~ Talon,
``Spin generalization of
the Calogero-Moser system and the Matrix KP equation'',
hepth/9411160}
\lref\gmmra{
A.~ Gorsky,~ A.~ Marshakov,~A.~ Mironov, A.~ Morozov,
``$N=2$ Supersymmetric QCD and Integrable Spin chains:
Rational Case $N_{f} < 2N_{c}$'', hep-th/9603140,  
ITEP/TH-6/96,
FIAN/TD-5/96}
\lref\kr{ I.~ Krichever, Funk. Anal. and Appl., {\bf 12}
(1978),  1, 76-78; {\bf 14} (1980), 282-290}
\lref\krz{I.~ Krichever, A.~ Zabrodin, ``Spin  
Generalization of
Ruijsenaars-Schneider model, Non-abelian Toda Chain and
Representations
of Sklyanin Algebra'', hep-th/9505039}
\lref\krphho{E.~D'Hoker, I. ~ Krichever, D.~ Phong,   
``The Effective
Prepotential of $N=2$ Supersymmetric $SU(N_c)$ Gauge  
Theories'',
hep-th/9609041\semi
I.~ Krichever and D.~  Phong,
``On the integrable geometry of soliton equations and N=2
supersymmetric
  gauge theories'', hep-th/9604199}
\lref\morozov{A.~ Marshakov, A.~ Mironov and A.~ Morozov,
``WDVV-like equations in $N=2$ SUSY Yang-Mills Theory'',
hep-th/9607109,
ITEP/TH-22/96, FIAN/TD-10/96}
\lref\matone{G.~ Bonelli and M.~  Matone,
``Nonperturbative Relations in $N=2$ SUSY Yang-Mills and WDVV
equation'', hep-th/9605090
}
\lref\fa{G.~ Faltings, ``A proof of Verlinde formula'',
J.Alg.Geom.{\bf 3}, (1994) }
\lref\beil{A. Beilinson, V. Drinfeld,
``Quantization of
Hitchin's Integrable System and Hecke Eigensheaves'',
{\rm A. Beilinson's lectures at IAS, fall 1994}}
\lref\be{D.~ Bernard, Nucl. Phys. {\bf B}303 (1988) 77,
{\bf B}309 (1988), 14}
\lref\calo{M.~ Bruschi, O.~ Ragnisco, ``Lax  
representations and
complete integrability for the periodic relativistic Toda  
lattice'',
Phys. Lett. A134, 365-370 (1989)\semi
V.I.~ Inozemtsev, ``The finite Toda lattices'', Comm.  
Math. Phys.
121, 629-638
(1989)}
\lref\c{F.~ Calogero, J.Math.Phys. {\bf 12} (1971) 419}
\lref\ch{I.~ Cherednik,
``Difference-elliptic operators and root systems'',  
hep-th/9410188}
\lref\eq{P.Etingof,
``Quantum integrable systems and
representations of Lie algebras'', hepth/9311132}
\lref\ek{P.Etingof, A.Kirillov, Jr., ``On the affine
analogue of Jack's and Macdonald's polynomials'', Yale  
preprint,
1994}
\lref\fv{G. Felder, A. Varchenko, ``Integral formula for the
solutions of the elliptic KZB  equations'', Int. Math.  
Res. Notices,
1995, vol. 5, pp. 222-233}
\lref\ga{R. Garnier, ``Sur une classe de system\`es  
differentiels
Abeli\'ens
d\'eduits de la theorie des \'equations lin\'eares'',  
Rend. del Circ.
Matematice Di Palermo, {\bf 43}, vol. 4 (1919)}
\lref\gau{M. Gaudin, Jour. Physique, {\bf 37} (1976),  
1087-1098}
\lref\iv{D.~ Ivanov,
`` Knizhnik-Zamolodchikov-Bernard equations on Riemann  
surfaces'',
hep-th/9410091}
\lref\kks{D.~ Kazhdan, B.~ Kostant and S.~ Sternberg,  
Comm. on Pure
and
Appl. Math., vol. {\bf XXXI}, (1978), 481-507}
\lref\lo{A.~ Losev, ``Coset construction and Bernard  
Equations'',
CERN-TH.6215/91}
\lref\op{M.~ Olshanetsky, A.~ Perelomov, Phys. Rep. {\bf  
71} (1981),
313}
\lref\m{J.~ Moser, Adv.Math. {\bf 16} (1975), 197-220; }

\lref\rkuper{S.N.M.~ Ruijsenaars, "Finite-Dimensionbal
Soliton Systems", in {\it Integrable
and Super-Integrable Systems}, Ed. B.A. Kupershmidt,
World Scientific, Singapore (1989)}
\lref\rs{S.N.M. Ruijsenaars,
H. Schneider, Ann. of Physics {\bf 170} (1986),
 370-405}
\lref\r{S. Ruijsenaars, CMP {\bf 110}  (1987), 191-213 }

\lref\s{B. Sutherland, Phys. Rev. {\bf A5} (1972), 1372-1376;}
\lref\sch{L. Schlesinger, `` \"Uber eien Klasse
von Differentialsystemen
beliebiger Ordnung mit festen kritischen Punkten'',
Journal f\"ur die reine und angewandte Mathematik,
Bd. CXLI (1912), pp. 96-145}
\lref\er{B. Enriquez, V. Rubtsov, ``Hitchin systems, higher
Gaudin operators and {\sl r}-matrices'', alg-geom/9503010}

\Title{ \vbox{\baselineskip12pt\hbox{hep-th/9609219}
\hbox{ITEP--TH--26/96}
\hbox{HUTP-96/A023}}}
{\vbox{
\centerline{Five Dimensional Gauge Theories}
\vskip 0.2cm
\centerline{and}
\vskip  0.2cm
\centerline{Relativistic Integrable Systems}}}
\bigskip
\bigskip
\centerline{Nikita Nekrasov\footnote{}{Junior Fellow,  
Harvard Society
of Fellows. On leave of  absence from:
Institute of Theoretical and Experimental Physics,
117259, Moscow, Russia}}
\vskip 1cm
\centerline{\it Lyman Laboratory of Physics, Harvard  
University,
Cambridge MA 02138}
\vskip 0.1cm
\centerline{nikita@string.harvard.edu}
\vskip 0.5cm
\centerline{\sl Talk given at the III International
Conference }
\centerline{\sl ``Conformal Field Theories and Integrable  
Models'',
Chernogolovka, June  23-30 1996}
\vskip 0.5cm
We propose a non-perturbative solution of $N=2$ supersymmetric
gauge theory in five dimensions compactified
on circle of a radius $R$.
We consider the cases of the pure gauge theory as well
as the theories with  matter in the fundamental and in  
the adjoint
representations. The pure theory as well as the one with
adjoint hypermultiplet give rise to the known
relativistic integrable systems with ${1\over{R}}$ playing the
r\^ole of the speed of light. The theory with adjoint  
hypermultiplet
exhibits some interesting finiteness properties.

\Date{September 1996}

\newsec{Introduction}
Recently there has been a tremendous progress in understanding
the non-perturbative behavior of supersymmetric gauge theories
in four dimensions. Namely, the exact low-energy  
effective action
was determined for a large class of $N=1$ and $N=2$ theories.
The main tool for such determination is the use of holomorphy
of various functions and the electric-magnetic duality,
constraining the geometry of the moduli space of scalars.

Of particular importance for our purposes is the solution
of the pure $N=2$ super-Yang-Mills theory originally
obtained by Seiberg and Witten \SeWi and reinterpreted later
by Gorsky et al. \gmmm in terms of integrable systems.
The integrable system, relevant to the
$N=2$ gauge theory with the massive adjoint hypermultiplet
was found by Donagi and Witten in \WitDonagi.

In this paper we address the following question.
Consider a five dimensional supersymmetric gauge theory.
Take the space-time manifold to be $M = X \times {\bf  
S}_{R}^{1}$,
where $X$ is a four-manifold and $R$ is the radius of the
circle $\bf S^{1}$. One can think of the theory
on $M$ as of the theory on $X$ with infinite number
of massive fields, according to Kaluza-Klein ideology.
It  seems that one can integrate out the massive fields  
and define
the effective action for the massless modes.
The effective theory has four-dimensional $N=2$
supersymmetry and the field content
of the abelian minimal theory. Our goal is
to determine its exact non-perturbative prepotential.

The low-energy  four dimensional theory contains the photon(s)
and therefore exhibits the usual electric-magnetic  
duality. This
leads to the fact that  the different regions of the  
moduli space
of vacua are  described by the different abelian  
theories, related
to each other by duality transformations. This is a  
common feature
of Coulomb branches of all $N=2$ gauge theories.

The novelty of our problem in comparison with the
conventional $N=2$ pure Yang-Mills theory is the
existence of a bigger group of discrete gauge transformations,
which are present in the Coulomb phase. For example,
in $SU(2)$ gauge theory in the $U(1)$ phase the
scalar $a$ related by $N=2$ susy to the photon is acted
on by the ${\IZ}_{2}$ transformation: $a \to -a$ which is
a remnant of the $SU(2)$ gauge invariance\footnote{1}{for  
the general
gauge group $G$ the scalars in the low energy theory live in
the Cartan subalgbera $\liet$ of the Lie algebra $\lieg$  
of $G$,
and the residual gauge invariance is the action of the
Weyl group $W$}. When the $N=2$ theory is obtained by the
compactification from five dimensions, the complex
scalar $\phi$ comes from the real scalar $\varphi$ of
the five dimensional theory and the fifth component  $A_{t}$
of the gauge field. Therefore, the residual gauge invariance
includes the shifts $a \to a + {{in}\over{R}}$,
coming from the gauge transformations of the form
$g(x,t) = \exp{ i t {n \over{R}}  
{\sigma}_{3}}$\footnote{2}{There is
a subtlety
in identifying the periodicity of $g(x,t)$ in $t$, for in  
the theory
with
adjoint fields only one can assume that the gauge  
transformations
are periodic up to the element of the center of the gauge  
group $G$
only. This
is discussed below.},
for $n \in {\IZ}$. This implies that the gauge invariant order
parameter
$$u \sim {\Tr} \phi^{2}  \sim  a^{2}$$ of the pure Yang-Mills
must be replaced by\footnote{3}{for the general
gauge group $G$ the residual gauge symmetry acting
on the scalars
is the affine Weyl group and the order parameters are
the characters of the group
element $\exp(i R \phi)$ in
the  irreducible representations of $G$ (there
are $r = {\rm rk}G$ independent characters)}
$$ U \sim {\Tr} \exp 2\pi R{\phi} \sim  {\rm cosh}  
\bigl(2\pi R {a}
\bigr)$$
(we are not careful with the numerical
coefficients in these formulae).
The low energy Lagrangian (more precisely,
its part containing no more then two derivatives and four
fermions) has the following form:
$$
{\CL} = {1\over{4\pi}} {\rm Im} \Biggl[
\int d^{4} {\theta} {{\p \CF}\over{\p A^{i}}} {\bar A}^{i} +
{\half} \int d^{2} {\theta} {{\p^{2} \CF}\over{\p A^{i}  
\p A^{j}}}
W_{\alpha}^{i} W^{\alpha, j}
\Biggr]
$$

The coupling constants $\tau_{ij} = {{\p^{2} \CF}\over{\p  
A^{i} \p
A^{j}}}$
non-trivially transform
under the electric-magnetic duality group  $Sp(2r,  
{\IZ})$. In the
rank one
(for  the gauge group $SU(2)$, say)  case the effective
gauge coupling constant
$$
\tau (A) = {{\p^{2} \CF}\over{\p A^{2}}} =
{{\vartheta}\over{2\pi}} + {{4\pi i }\over{g^{2}}}
$$
transforms under electric-magnetic duality group  
$SL_{2}({\IZ})$
as
$$
\tau \to {{m \tau + n}\over{r \tau + s}}
$$
One introduces a dual scalar $a_{D}$ (which is an $N=2$  
superpartner
of a dual photon): $a_{D} = {{\p \CF}\over{\p a}}$. The  
duality
transformations mix $a$ and $a_{D}$:
$$
\pmatrix{a\cr a_{D}\cr} \to \pmatrix{ r a_{D} + s a \cr m  
a_{D} + n
a\cr}
$$
Stated more accurately, $a$ and $a_{D}$ appear as the  
central charges
in the BPS
representations of the $N=2$ susy algebra. They depend
holomorphically
on the order parameter $U$ (or $u$). This dependence  
comes both from
the one-loop and instanton corrections.

Given the holomorphy and the duality properties of the  
solution we
look
 for a holomorphic section $\bigl( a(U),  a_{D}(U)
\bigr)$ of
an $SL_{2}({\IZ})$ bundle.

The solution of a pure Yang-Mills theory as presented in  
\SeWi\
makes  use of a family of elliptic curves $E_{u}$:
$$
i) \quad y^{2} = (x^{2}-\Lambda^{4}) ( x - u)
$$
or
$$
ii) \quad z + {{{\Lambda}^{4}}\over{z}} = p^{2} - 2u
$$
It has the form:
$$
\pmatrix{a(u) \cr a_{D}(u)\cr} =
\pmatrix{\int_{A} {\lambda}\cr \int_{B} {\lambda}\cr} $$
where $\lambda$ is a meromorphic one-differential
$$i) \quad \lambda = {{y dx}\over{\Lambda^{4} - x^{2}}}  
\qquad {\rm
or} \qquad
ii) \quad \lambda = p {{dz}\over{z}}
$$
Here $\Lambda$ is a dynamically generated mass scale in the
asymptotically
free theory.
This choice of a family of curves and the differential was
interpreted in \gmmm\ in terms of an integrable system:
make a change of variables:
$$ y = \Lambda^{2} p \sin {\varphi}, x = \Lambda^{2} \cos  
{\varphi},
z = \Lambda^{2} e^{i\varphi}$$
The
family of curves $E_{u}$ coincides with the  family of  
the levels
of a Hamiltonian of a periodic Toda chain:
$$
u = {{p^{2}}\over{2}} + \Lambda^{2} \cos{\varphi}
$$
while the differential $\lambda$ is nothing but the
canonical one-form $p d {\varphi}$.
The choice of contours of integration - the cycles $A$ and $B$
is determined by the study of the limit $u \to \infty$.
There $a \sim \sqrt{u}, a_{D} \sim  a \log ( u/\Lambda^{2} )$,
as suggested by the perturbatively exact beta-function.

We will show that the solution of five dimensional theory
can be obtained along the same lines with the periodic
Toda chain replaced by its relativistic analogue \rkuper :
$$
U ={1\over{R^{2}}} \cosh{Rp} \sqrt{ 1 + 2(\Lambda R)^2
\cos {\varphi} }
$$
For the $G = SU(N)$ case we obtain the corresponding
family of curves,  find the differential $\lambda$ and
compare it to the perturbative answer which we
obtain with the help of a  mathematical trick.

It is worth stressing that inspite of the fact that the  
additional
states we get
upon compactification are all massive with the masses $\sim
{1\over{R}}$
they do correct the metric on the space of vacua. The  
low-energy
effective action has only low momenta modes but it  
describes the
whole
moduli  space of vacua. As such, depending on the value  
of the scalar
$U$
the running  of the effective coupling stops at the
momentum (which runs inside the loop
and so is not restricted by the external momentum being  
small),
which can be very high in comparison to the external one.

The results of this paper extend the previous studies of the
five dimensional gauge theory \AFT. There, the perturbative
prepotential
for the five dimensional theory ($R = \infty$) was obtained
by the perturbative computation in the heterotic string  
and compared
with the predictions of the string duality (five dimensional
supersymmetry
arises in the compactifications of $M$-theory on a Calabi-Yau
threefold).
The prepotential in the compactifications of $M$-theory  
on CY is
given
by the intersection form of the CY (it is therefore
piece-wise cubic):
\eqn\prtprep{
\CF_{5d}  =  {1\over{6}} \sum_{i,j,k} C_{ijk} A^{i} A^{j}  
A^{k}}
Upon the further compactification on a circle
of the radius $R$ the $M$-theory becomes the $IIA$ string and
one expects the classical intersection ring of the CY be  
replaced
by its quantum cohomology which can be computed using mirror
symmetry\footnote{4}{I thank
C.Vafa for pointing this out  to me.}. Our proposal  
provides a {\it
local}
description of these mirror CY's. The remark on the  
relation between
the prepotential of the gauge theory in four dimensions  
and the
quantum cohomology of CY  doesn't explain the recently  
observed
phenomenon
that the prepotentials of the gauge theories
obey the WDVV equations \matone,  
\morozov\footnote{5}{This has been
explained to me by A. Morozov}.

\newsec{Five dimensional theory}

\subsec{Perturbative prepotential}

\subsubsec{One-loop evaluation}

The field content of a five dimensional supersymmetric
gauge theory consists of a gauge field $A_{m}$, a real
scalar $\varphi$
and a gluino $\lambda$
(all in the adjoint representation of the gauge group $G$).
We take the five-manifold $M$ to be the product
$M = X \times {\bf S}^{1}$ and denote
 the coordinates along $X$ as $x^{\mu}$,
while the coordinate along ${\bf S}^{1}$ will be denoted  
as $t$.

The partition function on $M$ in such a theory can be  
interpreted
as a partition function in the infinite-dimensional
supersymmetric quantum mechanics.

In order to see this explicitly let us analyze the
supersymmetry generators. One of the
 generators\footnote{6}{Upon dimensional reduction
down to four dimensions the theory becomes $N=2$
supersymmetric theory and the generator we are about
to consider corresponds to the scalar
supercharge in the twisted version - topological
theory} acts as follows (we change the notation
for the fermions - on a flat $X = {\IR}^{4}$ this doesn't  
matter,
on a general one
this corresponds to the field
content of a twisted theory. We also reintroduce an auxiliary
field, a self-dual two -form $H_{\mu\nu}^{+}$)  :
\eqn\susy{\eqalign{
        Q A_{\mu} = \psi_{\mu} \qquad & \qquad
Q \psi_{\mu} = F_{\mu t} + i D_{\mu}\varphi\cr
        Q (A_{t} + i \varphi) = 0 \qquad & \qquad
Q \chi_{\mu\nu}^{+} = H_{\mu\nu}^{+}\cr
       Q (A_{t} - i\varphi) = \eta \qquad & \qquad Q \eta  
= D_{t}
\varphi \cr
Q H_{\mu\nu}^{+} = D_{t} \chi_{\mu\nu}^{+} & + i [ \varphi,
\chi_{\mu\nu}^{+} ]\cr}}
This generator squares to the time translation together
with the space-like gauge transformation (they can be both
described as the result of the action of the operator
$\nabla_{t} = \p_{t} + A_{t} + i \varphi = D_{t} + i  
\varphi$).
It has the meaning of the equivariant derivative with  
respect to the
group
of loops in the group of
four-dimensional gauge transformations, extended
by the circle, rotating the loops.

The action of the theory is a $Q$ - commutator,
apart from the ``topological'' term:
\eqn\actn{
	S_{class} =  \int_{M} \theta \wedge {\Tr} F \wedge F
        + \{ Q , \int_{M}
\chi F^{+} - e_{0}^{2} \chi H + \psi_{\mu} (F_{\mu t} - i
D_{t}\varphi)
   \}
}
with $\theta$ being a background abelian gauge field
(it couples to the topological current ${\Tr} F \wedge F$).
We expand all fields in the Fourier series in $t$ and then
use the localization argument, which reduces
the path integral to the integral over the zero modes
along $t$ of the ratio of determinants one gets by expanding
the action around the $t$-independent field configuration
and treating the one-loop approximation.

In the low-energy limit the "would-be" potential
$(D_{t} \varphi)^{2}$ vanishes and we can fix the gauge
$$
A_{t} + i \varphi = \pmatrix{ a & 0\cr 0 & - a\cr}
$$
where $a$ is $t$-independent. This leaves
the freedom: $a \to -a$ and
$a \to a + {i}{{n}\over{R}}$, for $n \in \IZ$.

Now let us expand the action around the time-independent
configuration.

$$
S = S_{0} + S_{2}[A_{n}, \psi_{n}, \chi_{n}, \eta_{n},  
\varphi_{n} ]
$$
The supersymmetry guarantees the exactness of the one-loop
approximation.
The result of the evaluation of the integral
over the modes $A_{n}, \dots$ together
with the standard one-loop effective action
of the pure four-dimensional theory looks
like the standard low-energy action with the prepotential
$$
\CF^{\rm pert}_{5d}(a) = {{i}\over{8\pi^{3} R^{2}}}  
\sum_{\alpha}
{\rm  Li}_{3}\biggl(e^{2\pi R\langle \alpha, a \rangle}\biggr)
$$
($\Delta$ denotes the set of roots in $\lieg$)
which gives rise to the couplings:
\eqn\cplngs{
\tau_{\alpha} ={{i}\over{2\pi}} {\rm log} \Biggl(
{\rm sinh}^{2}  \bigl( \pi R \langle \alpha, a \rangle \bigr)
\Biggr).
}
This is the "quantum deformation" of the standard perturbative
prepotential of a pure $N=2$ theory:
$$
\CF^{\rm pert}_{4d}(a) = \sum_{\alpha \in \Delta}
{{i \langle a, \alpha \rangle^{2}}\over{8\pi}}
\log
{{ \langle a, \alpha \rangle^{2}}\over{\Lambda^{2}}}
$$
Indeed, in the limit $x \to 0$ the trilogarithm
${\rm Li}_{3}(e^{x})$ goes over to ${{x^{2}}\over{4}}  
\log x^{2} +
\dots$.

To understand \cplngs\  it is instructive to consider the  
$G=SU(2)$
case
where one has just one coupling $\tau(a)$ and rewrite it as:
\eqn\cplex{
\tau (a) = {{i}\over{2\pi}} {\log} \bigl(  
{{a^{2}}\over{\Lambda^{2}}}
 \bigr) +
\sum_{n=1}^{\infty} {{i}\over{\pi}} {\log} \bigl( a^{2}  + ( n
/R)^{2} \bigr)
}
One recognizes in the infinite sum the contribution of  
the excited
Kaluza-Klein states
(to avoid the confusion, this only takes into account the
electrically
charged states - the magnetically charged BPS strings are
non-perturbative and appear only in the final expression).
One also needs an ultra-violet regularization which gives  
rise to
$\Lambda$.

\sssec{Decompactification \quad  limit.} In the opposite  
limit  $x
\to \infty$ one has ${\rm Li}_{3} (e^{x}) \sim -  
{{x^{3}}\over{6}} +
$
exponentially suppressed terms. This limit reproduces  
\prtprep for
some
specific choice of the coefficients $C_{ijk}$.  In fact,  
in this
limit
all the non-perturbative corrections vanish
(as the contribution of the $n$ instantons is of the  
order of $\sim
e^{-{{nR}\over{g_{5}^{2}}}}$)
and one is left with
the purely perturbative
result. In fact, the large $x$ asymptotic of ${\rm  
Li}_{3}(e^{x})$
contains the quadratic as well
as the linear (in $x$) terms which are irrelevant in the  
large $R$
limit (one has
to divide the four-dimensional expression for the  
prepotential by
$2\pi R$ before
taking the limit).

One can also notice the non-trivial chamber structure of  
the moduli
space in
the decompactified theory.  It comes from the monodromy  
properties of
the
trilogarithm (see \hm\ for the
discussions of the similar effects in the context of
the heterotic string compactification). The answer for  
the five
dimensional
prepotential (the scalars $a_{i}$ become real) looks like:
\eqn\prepfd{
\CF_{5d} = {1\over{6}} \sum_{i,j} | a_{i} - a_{j} |^{3}}
There might be also a quadratic piece, corresponding to  
the finite
bare coupling in five dimensions \seifive.
The non-smoothness of $\CF_{5d}$ at the walls of the Weyl  
chamber has
to do with the restoration of non-abelian  gauge symmetry  
\wittrans.

\subsubsec{Instanton derivation}
{}From the mathematical point of view the
five dimensional theory is a way of study of
the geometry of moduli space of instantons on $X$.
More precisely, under some topological
restrictions the moduli space $\cal M$ is a spinor
Riemannian manifold and one may study
the index of Dirac operator $D$, acting
on spinors on $\CM$.

The index of $D$	is given by Atiyah-Singer	 
theorem:
$$
{\rm Ind}D = \int_{\CM} {\hat A}({\CM}) =
\int_{\CM} \prod_{i} {{x_{i}/2}\over{\sinh (x_{i}/2)}}$$
where the symmetric polynomials of the indeterminants  
$x_{i}$'s
are to be found from the Riemann-Roch-Grothendieck (or
Atiyah-Singer family index) theorem:\footnote{7}{In the  
general case
the $\hat A (X)$ must
be replaced by the Todd class $Td(X)$}
\eqn\atsn{
\sum_{i} e^{tx_{i}} =
\int_{X} {\hat A}(X)
{\Tr}_{V}\exp\bigl({{t\phi + t^{1/2}\psi +F}\over{2\pi  
i}}\bigr)
}
Here the fields $\phi$, $\psi$, $F$ are the scalar,
the one-form fermion and the curvature of the gauge field  
in the
twisted $N=2$ theory (which is often reffered to as the
Donaldson theory, as constructed by E.Witten \Witr). The  
trace is
taken in the adjoint representation $V$ (the details
are provided in \diss).
Formally one can pass from the expression for
$$
\sum_{i} x_{i}^{n}
$$
to the ${\hat A}({\CM})$.
One gets:
\eqn\gnfl{
{\hat A}({\CM}) = \exp \int_{X} \Biggl( {\CO}^{(4)}_{\CF} +
({\alpha} {\Tr} R^{2} + {\beta} {\Tr}R{\tilde R})
{\CO}^{(0)}_{\CP} \Biggr)
}
where ${\CO}^{(i)}_{A}$ denotes the $i$-observable,
constructed by means of the standard descend equations \Witr\
out of the gauge invariant functional $A(\phi)$. The  
coefficients
$\alpha$ and $\beta$
in front of the densities ${\Tr} R \wedge R$ and
${\Tr} R \wedge {\tilde R}$ are such that
$$
\int_{X} {\alpha} {\Tr} R^{2} + {\beta} {\Tr}R {\tilde R}  
= \biggl(
{{{\chi}+{\sigma}}\over{4}} \biggr)
$$
where
$\chi$ and $\sigma$ denote Euler characteristics
and  signature of the manifold $X$ respectively.

The passage from  \atsn\ to the ${\hat A}$-genus is very
similar to the evaluation of the one-loop
diagrams. It  leads to the following expressions
for the functions $\CF$ and $\CP$ entering \gnfl\
(we write the value of the functionals on the $\phi = a  
\in \liet$
and we introduce $R$ to keep track of dimensions):
\eqn\prep{\eqalign{
	\CF (a) + \CF^{\rm pert}(a)_{\rm pure YM}\qquad = &
	\quad  {{i}\over{8\pi ^{3}R^{2}}} \sum_{\alpha  
\in \Delta}
	{\rm Li}_{3} \bigl(e^{2\pi R\langle a, \alpha  
\rangle}\bigr) \cr
	\CP (a) \qquad = &\qquad {\rm  \log}  
\prod_{\alpha > 0}
	{{\pi R\langle a, \alpha \rangle}\over{{\rm  
\sinh}\bigl(
	{\pi R\langle a, \alpha \rangle \bigr)}}} \cr}}
(In other words, $\p^{2} \CF / \p x^{2} \sim -  
\log(1-e^{-x}) -
\log (1 - e^{x}) + \log x^{2}$, for $x = \langle \alpha ,  
2\pi a
\rangle$).
In the formula \prep\ we use the expression for the
perturbative prepotential of the pure $N=2$ theory:
$$
\CF^{\rm pert}(a)_{\rm pure YM} =
\sum_{\alpha \in \Delta}
{{i \langle a, \alpha \rangle^{2}}\over{8\pi}}
{\log}
{{ \langle a, \alpha \rangle^{2}}\over{\Lambda^{2}}}
$$

For example, for $G=SU(2)$ the perturbatively exact   
coupling is
given by
the following  expression:
\eqn\bet{\tau (a) = {i\over{2\pi}} {\rm log}  
\bigl({1\over{(\Lambda
R)^{2}}}{\rm sinh}^{2}({{aR}\over{2}})\bigr) }
One easily recognizes the presence of the group-like  
Vandermonde
determinant
in this formula (compare with  the Lie-algebraic one in the
perturbative
four-dimensional formulae).

\subsec{Non-perturbative answer}

\subsubsec{The case $G = SU(2)$}
Recall the answer for the pure four dimensional theory,  
as presented
in \gmmm. Consider  the periodic Toda chain (for $G = SU(2)$)
$$
h(p,q) = {{p^{2}}\over{2}} + \Lambda^{2} \cosh(q)
$$
There is a family of elliptic curves $E_{u}$:
$$
u = h(p,q)
$$
associated with this system. Here
 $p$ and $q$ are allowed to vary in the complex region.
The claim of \SeWi,\gmmm\ is that the modular parameter
$\tau (u)$
of the curve $E_{u}$ in the family coincides with the
gauge coupling in the low-energy abelian theory. The fact
that $\tau (u)$ is defined up to the modular transformation
corresponds to the electric-magnetic duality
phenomenon in the abelian gauge theory.

In our case the low-energy four dimensional
theory exhibits the electric-magnetic duality as well
(in the purely five-dimensional theory the vector
multiplet gets mapped to the tensor multiplet
under the duality transformation,
but upon compactification on the circle the
tensor multiplet becomes equivalent to the
vector multiplet as well)
so it is natural to look for a family
of elliptic curves encoding
the answer for the non-perturbative prepotential.

We claim that this family is a relativistic  
generalization \rkuper
of the
Toda chain:
$$
H(p,q) = {1\over{R^{2}}}\cosh( Rp) \sqrt{1 +
2
(\Lambda R)^{2}
\cosh(q)}
$$
In the limit $\Lambda R \to 0$ this Hamiltonian
reduces to the standard one (in the notations of
Ruijsenaars $1/R$ corresponds to the speed of light).

The family of the curves $\CE_{U}$
defined as $U = H(p,q)$
yields the correct asymptotics of the coupling constant  
$\tau(U)$.
It is instructive to study the periods
themselves ($\zeta = \Lambda R $):
\eqn\prds{\eqalign{
{\vec A} \qquad =& \qquad \oint_{\vec \gamma} pdq \cr
{{\p \vec A}\over{\p U}} \qquad =& \qquad
\oint_{\vec \gamma} {{R  
dq}\over{\sqrt{R^{4}U^{2}-1-2{\zeta}^{2}
\cosh(q)}}}\cr}}
Using
$$
U={1\over{R^{2}}} \cosh{\alpha}, \qquad
\nu_{5} = 2{{\zeta^{2}}\over{\sinh^{2}{\alpha}}}$$
\prds\  can be rewritten as:
\eqn\prdsii{
{{\p \vec A}\over{\p \alpha}} \qquad = \qquad {1\over{R}}
\oint_{\vec \gamma}  {{dq}\over{\sqrt{1 - {\nu}_{5}  
\cosh(q)}}}}
Let us  compare \prds\ with the periods $\vec a$ of the
pure four dimensional theory, where for:
$$
u = {{{\aleph}^{2}}\over{2}}, \qquad
\nu_{4} = {{{\Lambda}^{2}}\over{u}}$$
one has:
\eqn\swprds{\eqalign{
{{\p \vec a}\over{\p \aleph}}\qquad =& \qquad  
{1\over{2\pi \Lambda}}
\oint_{\vec \gamma} {{dq}\over{\sqrt{1 - \nu_{4}  
\cosh(q)}}}\cr}}

This implies that the curves $E_{u}$ and $\CE_{U}$
are isomorphic
for $\nu_{4} = \nu_{5}$, i.e.
for
$$
u = R^{2}U^{2} - {1\over{R^{2}}}
$$
In particular the asymptotic $\tau (u) \sim {\rm log}
({u\over{\Lambda^{2}}})$ of
the curve $E_{u}$ gets mapped to $\tau (U) \sim  {\rm log}
({{R^{4}U^{2}-1}\over{R^{2}\Lambda^{2}}})$ for the curve  
$\CE_{U}$.
Together with the asymptotic $U \sim {1\over{R^{2}} }  
{\rm cosh}
(2\pi AR)$
this yields \bet.

The elliptic curve $E_{u}$ degenerates for ${\nu}_{4}= \pm 1,
\infty$.
This corresponds to
$$
U = \pm {1\over{R^{2}}}\sqrt{1 \pm \zeta^{2}} ,\infty
$$
Notice that in the limit $R \to \infty$ all singular points
different from $U = \infty$ collide at $U = 0$, which  
corresponds
to the restoration of the non-abelian gauge symmetry.  
This phenomenon
has been noticed by several authors (e.g. \AFT,  
\wittrans). Also
notice that we have five singular points as opposed
to the notorious three points of the four-dimensional
theory. This relies on the fact that
in the absence of the fields in the fundamental
representation the low-energy theory has a ${\IZ}_{2}$
symmetry, which maps $U$ to $- U$. This symmetry has the  
following
origin. The order parameter $U$ is nothing
but the trace in the fundamental representation of the  
"Wilson loop"
$g = P\exp\oint (A_{t} + i\varphi)$. If all fields under
consideration are in the adjoint
representation of the group $G$, then the multiplication  
of the
Wilson loop $g$
by
the element of the center of the gauge group
does not change physics. Therefore the true moduli space  
of the
theory
is the quotient of the $U$ plane by the symmetry
$U \to - U$. This symmetry is broken as long as
one has a fundamental matter.

\subsubsec{The general case: $G= SU(N)$}

In the rank $N-1$ case the integrable system behind the
four dimensional theory is known to be the
periodic Toda chain. It is an integrable system on the
$2(N-1)$ complex dimensional phase space with the
canonical coordinates
$(p_{i},q_{i})$, $i = 1, \dots , N$ which obey the constraint:
$$
\sum_{i} q_{i} = \sum_{i} p_{i} = 0
$$
As any integrable system is has $N-1$ integral in involution,
i.e. the set of functions $I_{2}, \dots, I_{N}$
of $p$ and $q$ such that they Poisson-commute and
are functionally independent. The simplest way to
see these integrals of motion is through
the Lax operator. It is a $N \times N$ matrix $L$, whose
entries depend on $p, q$ and also on the auxiliary
parameter $z$:
\eqn\todalx{\eqalign{
& L_{ij}(z)  = p_{i} \delta_{ij} +
\delta_{i, j-1} +
\delta_{i, j+1} {\Lambda^{2}} e^{q_{i} - q_{i-1}} \cr
& - \Lambda^{N} z \delta_{i,N} \delta_{j,1} +
\Lambda^{2-N}  z^{-1} e^{q_{1} - q_{N}}
\delta_{i,1}\delta_{j,N}\cr}}
By definition,
the generating function of the integrals of motion
of the Toda lattice is the characteristic polynomial
\eqn\todai{
{\rm det}( L(z) - w) = \sum_{k=0}^{N} w^{k}I_{k}  +
 (-\Lambda)^{N}z +  \Lambda^{N}z^{-1} }

The action-angle variables of the system play
the vital role in the solution of the supersymmetric
gauge theory.
Namely, the action variables become the central charges  
of the susy
algebra $a^{i}$
which encode the behavior of couplings $\tau_{ij}$.  The angle
variables
are hidden in the four-dimensional
story but become visible (as moduli)
as long as one compactifies one dimension --
they become the vacuum expectation values of the
fourth component of the
gauge field and of the scalar dual to the three dimensional
photon  (see \witseiii).

The generalization of this story  to the theory on  
${\IR}^{4} \times
{\bf S}^{1}$
makes use of the Lax operator for the relativistic periodic
Toda chain (see, for example \rkuper, \calo):
\eqn\reltodlx{\eqalign{
L_{ij} & = e^{Rp_{i}} f _{i} (l_{ij} + b_{ij})\cr
l_{ij}  &=  \delta_{i,j+1} (1 + \zeta^{N} z) \xi_{i} -
\delta_{i,1}\delta_{j,N} (1 + \zeta^{-N} z^{-1}) \xi_{1} \cr
b_{ij}  &=   \Biggl[
\eqalign{-(i\zeta)^{N} ,& \quad i \leq j-1\cr
1, &\quad i > j-1 \cr}\cr
f_{i}^{2}  & = (1 - \zeta^{2} e^{q_{i+1}-q_{i}} )(1 -  
\zeta^{2}
e^{q_{i}-q_{i-1}} )\cr
\xi_{i}^{-1} & = 1 - \zeta^{-2} e^{q_{i-1} - q_{i}}
\cr}}
with $q_{N+1} = q_{1}, q_{0} = q_{N-1}$.

The spectral curve ${\rm det}(L(z) -w) =0 $ can be  
constructed using
the results
of \rkuper. One has:
\eqn\relsp{\eqalign{
& {\rm det}( L(z) -w ) = \sum_{l=0}^{N}   (-w)^{N-l}   
{\sigma}_{l}\cr
& = (-w)^{N} + \sum_{l=1}^{N-1} (1 +\zeta^{N} z)^{l-1}  
(-w)^{N-l}
S_{l} \cr
& \qquad +
( 1 + \zeta^{N} z)^{N-1} (1 + \zeta^{N} z^{-1} )\cr}}
where ${\sigma}_{l}$ is the $l$'th symmetric function of  
$L(z)$ and
$S_{l}$ are the $z$-independent coefficients.
After the change of variables
\eqn\chngs{
y = - {{1 + \zeta^{N} z}\over{w}}, \quad \chi = z y^{-N/2} }
the curve acquires the form
\eqn\crviii{
y^{-N/2} \CP(y)  = \zeta^{N} ({\chi} + {\chi}^{-1})}
with $\CP$ denoting the  $N$'th degree polynomial (which  
starts with
$1$ and
ends with $y^{N}$)
and the differential is
$$
\lambda = {1\over{2\pi i}} {\log} (y )  {{d  
{\chi}}\over{\chi}}
$$
The proposal passes through  a number of checks. The  
first one is
the comparison to the known solutions of the $N=2$  
super-Yang-Mills
theory with the gauge group $G = SU(N)$ \argfar.
The solution makes use of the family
of hyperelliptic
curves
\eqn\argcrv{y^{2} = P(x)^{2} -
{\Lambda}^{2N}}
where
$P(x) = {\rm det}(x - \phi)$ is the characteristic
polynomial of the Higgs field. In the Coulomb
phase the Higgs field can be diagonalized:
$\phi = {\rm diag}(a_{1}, \dots, a_{N})$.
$$
{\rm det}(\phi -x) = \prod_{i} (a_{i} - x)
$$
The five-dimensional gauge theory with
the gauge group $G$ can be understood as a four
dimensional theory with the gauge group $\hat G$ -
the extended loop group. The analogue of
\argcrv for $\hat G$ is easy to find: replace the
Higgs field $\phi(x)$ by
the first order differential operator on the circle:
$$
\phi \to
\p_{t} + A_{t} + i\varphi
$$
The determinant
${\rm det}(\p_{t} + A_{t} + i\varphi - x)$
can be conveniently defined
using the zeta-regularization with the result:
$$
{\rm det}(\p_{t} + A_{t} + i\varphi - x) =
\prod_{i} {1\over{R}} {\rm sinh}(2\pi Ra_{i} - x)
$$
Therefore the curve \argfar\ assumes the form:
\eqn\newcrv{y^{2} =
\prod_{i}{1\over{R^{2}}}  {\rm sinh}^{2}(2\pi Ra_{i} - x) -
{\Lambda}^{2N}}

The convenient  curve
\gmmm
\eqn\zfrm{z + {{\Lambda^{2N}}\over{z}} = P(x)}
goes over to
\eqn\newzfrm{z + {{\Lambda^{2N}}\over{z}} =
\prod_{i}\biggl(
{
{
{\rm sinh}
( 2\pi Ra_{i} - x)}\over{R}}
\biggr) }
Introduce the variables $y = e^{2 x}$, $\chi =
{{z}\over{\Lambda^{N}}}$. Then, thanks
to $\sum_{i} a_{i} = 0$ we rewrite \newzfrm  as:
\eqn\nnewzfrm{
{\zeta}^{N} \bigl( {\chi} + {{1}\over{\chi}} \bigr) = y^{-N/2}
\CP_{N} (y)}
with
\eqn\pp{\CP_{N}(y) = 1 + I_{1} y + \dots + I_{N-1}y^{N-1}  
+ y^{N}}
which coincides with the spectral
curve \crviii
of the relativistic periodic Toda system. We observe  the
${\IZ}_{N}$ degeneracy of the
five-dimensional moduli space compared  to the  
four-dimensional one :
under the transformations
\eqn\cent{
y \to e^{{2\pi i k}\over{N}} y, \quad I_{l} \to e^{-{{2\pi i
kl}\over{N}}} I_{l},
\quad z \to (-)^{k} z}
the curve and the differentials $\p \lambda / \p I_{l}$
are not changed. These transformations correspond
to the multiplications of the Wilson loop
$P\exp\oint (A_{t} + i\varphi)$ by the elements of the  
center of
$SU(N)$.

\subsec{$N_{f} > 0$}

The case with the fundamental matter can, in principle, be
treated with the help of integrable systems, although we were
not able to identify them (see \gmmra, though). Instead  
for the sake
of brevity we
present the family of curves with the differentials using the
remarks at the end of the previous section.

The family of curves describing  the $N=2$ theory
with $N_{f}$ hypermulitplets in the fundamental  
representation of
$G = SU(N_{c})$ gauge group (see also
\krphho for the more general case) is known.

\eqn\crvmat{\eqalign{
z + {{ \Lambda^{2N_{c}- N_{f}} P_{N_{f}}(x)}\over{z}}  = &
P_{N_{c}}(x)\cr
P_{N_{f}} (x) = & \prod_{k=1}^{N_{f}}  ( x + m_{k} ) \cr
P_{N_{c}} (x) = & {\rm det}(x - \phi) \cr}}

In five dimensions the mass of a hypermultiplet is a real  
scalar
which can be thought of the scalar component of the  
vector multiplet.
Upon compactification on the circle of the
radius $R$
it becomes a complex scalar,
whose imaginary part lives on the circle
of the radius $1\over{R}$, i.e. the physics must be  
invariant under
$m_{i} \to m_{i } + {{i n_{i}}\over{R}}$ for $n_{i} \in \IZ$.
This suggests the replacements:
\eqn\ncrvmat{\eqalign{
P_{N_{f}} (x) = & \prod_{k=1}^{N_{f}} {1\over{R}} {\rm  
sinh}( x +
2\pi Rm_{k})\cr
P_{N_{c}} (x) = & \prod_{i=1}^{N_{c}} {1\over{R}} {\rm  
sinh}(x - 2\pi
Ra_{i}) \cr}}
Introducing once again $y = e^{2x}$ we write the new  
family of curves
as:
\eqn\newcrvs{
z + {{\CP_{N_{f}}(y)}\over{z y^{N_{f}/2}}}  =
y^{-N_{c}/2}\CP_{N_{c}}(y)}
with $\CP_{\ldots}$ again denote the relevant polynomials.

\subsec{Adjoint matter}
The theory with the massive adjoint hypermultiplet is  
peculiar in the
following
sense: it is ultra-violet finite in four-dimensions and seems
to be ultra-violet finite in five dimensions too. The  
perturbative
coupling  behaves like ($m$ is the mass of the
hypermultiplet, $G = SU(2)$):
\eqn\prtbh{
\tau^{eff} \sim \tau^{bare}+ {i\over{2\pi }} {\rm log} {{{\rm
sinh}^{2}(Ra)}\over{{\rm sinh}(Ra + 2\pi R m )
{\rm sinh}(Ra - 2\pi Rm)}}
}
and therefore stops running at  sufficiently
large   $a$. Assuming this finiteness
we introduce the microscopic coupling $\tau$ (the issue of the
relation of the five dimensional real coupling and the  
complex $\tau$
is discussed below). Now we look for the integrable  
system, which
has to be defined with the help of the elliptic curve  
$E_{\tau}$,
reduces at $R = 0$
to the elliptic Calogero-Moser system (as it describes
the four dimensional story \WitDonagi) and also reduces to the
relativistic Toda system in when hypermultiplet becomes  
heavy enough.

Fortunately, the collection of relativistic integrable systems
contains the one
which is the natural candidate for the theory with adjoint
hypermultiplet.
It is the elliptic Ruijsenaars-Schneider model
(or Relativistic Calogero-Moser system) \rs.

It has the Lax operator:
\eqn\elllx{
L_{ij} = e^{Rp_{i}} f_{i}(q)  {{{\sigma}( q_{ij} + z )  
{\sigma}
(iRm)}\over
{{\sigma}( q_{ij} + iRm ) {\sigma} (z)}}
}
where
\eqn\ffct{
f_{i}(q) ^{2} = \prod_{k \neq i}  \bigl( 1 -
{{{\wp}(q_{ik})}\over{{\wp}(iRm)}} \bigr),
}
$q_{ij} = q_{i} - q_{j}$, $\sigma$ is the Weierstrass  
sigma function
and
$\wp = - ( \log \sigma)^{\prime\prime}$.
For example, the Hamiltonian which replaces
$$
H^{non-rel} = \sum_{i} \half p_{i}^{2} + m^{2} \sum_{i  
\neq j} {\wp}
(q_{ij} )
$$
has the form
$$
H^{rel} = {1\over{R^{2}} } \sum_{i} {\rm cosh}(Rp_{i})  
f_{i} (q)
$$
Let us  show that the parameter $m$ in the formulae  
\elllx, \ffct\
indeed
corresponds to the mass of the hypermultiplet. More  
precisely, we
shall show
that the symplectic form $\sum_{i} d p_{i} \wedge d q^{i}$
has a non-trivial period, proportional to $m$ (this makes
contact with the ideology of \SeWi).\footnote{8}{This  
statement
is local in the $m$ space. As one goes around the cycle  
$m$ gets
shifted
and so does the period. This has to do with the monodromy
representation
of the
$\pi_{1}$ of $m$ space in $H_{2}({\CM}, {\IZ})$. } To  
simplify things
we
restrict ourselves with $G=SU(2)$ case and
consider  the spectral curve
\eqn\ellcrv{
U = {\rm cosh}(Rp)  \sqrt{ 1 -  
{{{\wp}(q)}\over{{\wp}(iRm)}} } }
The symplectic form $dp \wedge dq$ can be rewritten
as:
\eqn\smpl{
\omega = dp \wedge dq  = {
{dy \wedge dq}\over{2R \sqrt{(y - {\wp } (iRm) )(y -  
{\wp}(q))}}
}}
where $y = {\wp }(iRm) + {\rm cosh}^{2}(Rp) ({\wp}(q)  
-{\wp}(iRm)) $.
Now we describe a non-contractible
two-cycle $\Gamma$ and compute the integral
$$
\int_{\Gamma} \omega
$$
In the $y$-plane for the fixed $q$ there is a cut between  
the points
${\wp}(q)$ and ${\wp}(iRm)$.
There is a one-cycle $C_{q}$ which goes around the cut.  
The period
$$
\oint_{C_{q}} \alpha(q)
$$
of the one-differential
$$
\alpha (q) = { {dy}\over{\sqrt{(y - {\wp } (iRm) )(y -  
{\wp}(q))}}}
$$
equals $2\pi i$. The cycle $C_{q}$ vanishes when ${\wp}(q)
={\wp}(iRm)$.
Therefore there is a two-cycle $\Gamma$ (a two-sphere)
whose projection onto the
elliptic curve $E_{\tau}$ (where $q$ lives) is the contour
going from $q = - iRm$ to $q = + iRm$ and
the fiber over a point $q$  is $C_{q}$. Clearly,
\eqn\prds{
{1\over{2\pi i}} \int_{\Gamma} \omega = {1\over{2R}}
\int_{-iRm}^{+iRm} dq =
i m \quad
\bigl(+ {1\over{2R}} (n_{1} + n_{2} {\tau}) \bigr)
}
(we see that the periods are half-integral. This has to  
do with
the center of $SU(2)$)

One can proceed  similarly in the general case $N>2$.  
Notice that for
the two
degenerations of the elliptic Ruijsenaars-Schneider  
model, namely,
for the
trigonometric limit ($\tau \to \infty$) and for
the non-relativistic limit ($R \to 0$) the linear  
dependence of the
cohomology
class of the symplectic form $\omega$ follows from a  
holomorphic
variant of the
Duistermaat-Heckman theorem, as in those cases there is a
description of the models via the Hamiltonian reduction \gnru,
\gnell. Recently
the elliptic Ruijsenaars model has been given the Hamiltonian
reduction description \frolov.

\sssec{A \quad  bizarre \quad symmetry}. Notice that the  
solution of
the
model with the adjoint hypermultiplet has the following  
symmetry:
\eqn\bzsm{
m \to m + {i\over{2R}} ( n_{1} + n_{2} \tau ), \qquad  
n_{1,2} \in
\IZ}
The first symmetry ($n_{1}$) reflects the presence of the
tower of Kaluza-Klein states in
the four dimensional  theory. To understand the second one we
have to recall the relation between the couplings in four  
and five
dimensions.
In five dimensions $1\over{g_{5}^{2}}$ has the dimension  
of the mass
and enters the vector multiplet whose vector component gauges
the topological current ${\Tr}F \wedge F$. Again, upon
compactification
down to four dimensions on a circle it becomes a complex
scalar $\tau$ which is conventionally written as:
$$
\tau = {{4\pi i R}\over{g_{5}^{2}}} + {{\theta}\over{2\pi}}
$$
where $\theta$ is the integral of the corresponding gauge  
field along
the circle $\bf S^{1}$.
So, if  the bare coupling in five dimensions is finite  
then the
theory
must have a symmetry:
$$
m \to m + {{n}\over{g_{5}^{2}}}, \quad n \in \IZ
$$
already in the decompactification
limit.  We suspect that the presence
of the tower of states here has to do with
the tensionless strings in
six dimensions.
Upon the compactification on a small circle
the six dimensional string gives rise to the
tower of particles in five dimensions.
Indeed it was suggested by E. Witten in \witremarks\
that the $S$-duality of $N=4$ SYM in four dimensions
is best understood as the consequence of the existence of the
self-dual
non-critical  string in six dimensions. In five dimensions
the theory with the massless adjoint hypermultiplet gives rise
to $N=4$ SYM in four dimensions.  The moduli space seems to be
locally flat (as it is in four dimensions), the only  
difference being
that
the scalar $a$ lives on a circle.  In other words, the  
vector moduli
space
is ${\bf T}_{\IC}/W$, where ${\bf T}_{\IC}$ is the
complexification of the Cartan subgroup of the gauge  
group $G$.

On the other hand, if we are to have a finite coupling in  
the limit
$R \to 0$
then $g_{5}^{2}$ better be of the order of $R$ and  
therefore the
symmetry
\bzsm becomes vacuous.

The proposed solution  identifies the periods of the  
differential
$\lambda = p dz$ (the action variables) with the susy central
charges.
The subtlety here is that
the spectral curve for $N >2$ has a genus $N(N-1)/2$ as  
opposed to
the
non-relativistic case. The problem of identification
of the relevant periods has been recently solved \krzabr.

\sssec{Flow \quad  to \quad  the \quad pure \quad gauge \quad
theory}. One expects to
recover the minimal theory in the limit where the adjoint
hypermultiplet becomes very heavy.
It is clear, though, that the microscopic coupling $\tau$  
must be
tuned in such a way that
the limit $m \to \infty$ becomes possible.

Historically, the relativistic Calogero-Moser model was  
discovered
before the
corresponding Toda system \rs. The trick producing the  
periodic
Toda system out of the elliptic Calogero-Moser system is  
precisely
the flow we need. It is completely analogous to the  
non-relativistic
case.
One shifts the $q_{i}$ variables as:
$$
q_{j}  \to q_{j} - {j\over{N}} {\tau}
$$
After that one takes $\tau \to i \infty$, keeping
\eqn\ruiren{
\Lambda R = \exp
\bigl( 2\pi R {\bar m}  + 2\pi i {{\tau}\over{N}} \bigr)
}
(with $\bar m = m \cdot {\rm sgn}({\Re} m)$)
finite \rkuper.  This renormalization is also suggested by
\prtbh.  One has to be careful comparing \ruiren\  to the  
$R=0$ case
\WitDonagi, since
it corresponds to the hypermultiplet which is much  
heavier then the
 Kaluza-Klein modes ($m >> {1\over{R}}$)
and therefore the RG flow looks different from
the conventional one (which corresponds to ${1\over{R}}  
>> m >> a$).
 Given the
existence of all four integrable systems (Toda,  
Calogero-Moser and
their relativistic analogues)
together with their limiting properties one can  
reconstruct the
whole picture of flows. We plan to  return to this issue  
elsewhere.

\newsec{Future directions and conclusions}

Among the immediate future generalizations one can recognize
the solution of the
six-dimensional gauge theory. The corresponding Hamiltonians
are elliptic functions on the rapidities $p$
(this is in the slight contradiction with  the
statements of \martinecsix ). The corresponding integrable
model has not been known before and  can be constructed  
along the
lines of \dual,
as dual to the known elliptic systems. The perturbative  
prepotential
for this model has been computed in \diss. It corresponds  
to the
elliptic genus of the moduli
space of instantons (which is a generalization
of the $\hat A$ -genus). An interesting step towards
the actual computation of the elliptic genera has been made
in \dvvm.
The compactifications of
the six-dimensional theory has been recently
studied in a several contexts.  In particular, the sums of the
trilogarithms
appeared in \hm. The low-energy effective action
for the six-dimensional tensionless string
compactified on a two-torus was studied in \ganor.

It is also
interesting to compare our results with those of \witseiii.
Consider the five dimensional theory compactified on a  
two-torus
down to three dimensions. The moduli space is now
$4r$ dimensional for $r = {\rm rk}G$ and has a hyperk\"ahler
structure. This structure is parameterized by the
geometry of the two-torus. For
example for a rectangular torus it
depends on two radii $R_{4},R_{5}$ and in the
limit when one of the radii goes to zero we get the  
Atiyah-Hitchin's
monopole space ${\widetilde \CM_{2}^{0}}$ \witseiii. In  
the opposite
case where one of the radii goes to infinity we
expect to see the relativistic integrable system.

It is even more interesting to start with the six dimensional
theory.
Let us imagine that we have an $SU(2)$ (what are
we saying is possible to extend for a generic gauge group
$G$, but we restrict ourselves with $G=SU(2)$. This is also
relevant to $M$-theory compactification on $K3 \times  
{\IR}^{7}$)
$N=1$ supersymmetric gauge
theory on ${\bf T}^{3} \times {\IR}^{3}$. We take the metric
$G_{ab}d\varphi^{a}d\varphi^{b}$
on ${\bf T}^{3}$ to be flat to preserve supersymmetry.

The classical moduli space $\CM$ of the effective three
dimensional theory is a copy of an orbifold limit
of $K3$ manifold (this has been recently
noticed by the several authors \witlec, \seibbr). Indeed,  
the scalars
in the effective
theory are the Wilson loops around three cycles in ${\bf   
T}^{3}$
and the scalar, dual to the three dimensional photon.
The Wilson loops are the commuting elements of $G$ and
can be simultaneously diagonalized, i.e. they define
three elements of $U(1)$. They are defined uniquely
up to the transformation in the Weyl group, which
also acts on the scalar $\sigma$, dual to the photon.
The scalar $\sigma$ lives actually on the circle of the
circumference $2\pi$. Thus we get the manifold
of classical vacua, isomorphic to ${\bf T}^{4}/{\IZ}_{2}$.

The classical metric on the moduli space is flat
with the orbifold singularities at the fixed points
of ${\IZ}_{2}$ action. It seems that the quantum corrections
make the metric smooth at this points. The $N=4$ $3d$  
supersymmetry
guarantees that it should be hyper-K\"ahler.

Our solutions probe the different regions in the $K3$  
moduli space,
where
(for four- and five-dimensional theories) $K3$ looks  
non-compact in a
number of directions.

Our results in the limit $R \to \infty$ reproduce the  
formulas of
\AFT.
It would be interesting to
find the prepotentials (as the functions of $R$)
for  other examples considered in
\wittrans, \seifive,\morrseib, \dkv.

Our results will be used in the derivation of the four  
dimensional
analogues of Verlinde formulas \ver\. They count the number of
holomorphic blocks in the theories, which were
constructed and studied in \avatar,\cocycle, \diss, \clash.
\newsec{Acknowledgements}

I am grateful to A.~ Gerasimov, A.~ Gorsky, A.~ Losev,
A.~ Marshakov, A.~ Mironov, A.~ Mikhailov,
A.~ Morozov,
G.~ Moore, A.~ Rosly, S.~ Shatashvili and C.~ Vafa
for fruitful discussions.

The research was supported by the Harvard Society of Fellows
and in part by NSF via the grant  PHY-92-18167, and by  
RFFI via the
grants 96-02-18046, 96-15-96455.

\newsec{Appendix. Supersymmetric Quantum Mechanics on the  
Moduli
Space}

In this section we discuss the five-dimensional theory on  
a manifold
$X \times S^{1}$ in the limit where $X$ is effectively  
very small.
The theory reduces to the quantum mechanics whose target  
space is
the moduli space of instantons on $X$.

Recall that the action of the theory has the form
\eqn\actnii{\eqalign{
S = \int_{X \times {\bf S}^{1}} \theta \wedge
{\Tr} (F \wedge F) + &
\{ Q , \alpha_{m} R_{m} + \alpha_{a} R_{a} + \alpha_{k}  
R_{k} \} \cr
R_{a} = \eta D_{t} \varphi \qquad & \qquad
R_{k} = \psi^{\mu}  (F_{\mu t} - iD_{\mu} \varphi) \cr
R_{m} = i \chi^{\mu \nu} & (F_{\mu\nu}^{+} - e_{0}^{2}
H_{\mu\nu}^{+}) \cr
}}
where $\theta$ is the background gauge field
(it is  in the vector multiplet,
whose scalar component is nothing
but the five dimensional
bare coupling $1\over{e_{0}^{2}}$).

If all the observables are $Q$-closed we are free to vary the
parameters
$\alpha_{m}, \alpha_{k}, \alpha_{a}$ as long as appropriate
non-degeneracy conditions are fulfilled. The answer will  
remain
intact.

\sssec{Weak  \quad coupling \quad limit.}
By the weak coupling we understand here the limit
where $\alpha_{m} \to \infty$, $e_{0}^{2} \to 0$. We also
can assume that $\alpha_{a} \to 0$.
As $\alpha_{m}$ multiplies $H^{\mu\nu}F_{\mu\nu}^{+}$ the  
limit
$\alpha_{m} \to \infty$
constrains the space\footnote{9}{We refer to $X$ as to the
space}-like part $A_{\mu}$  of the gauge
field to be anti-self-dual:
$$
F_{\mu\nu}^{+} = 0 \qquad (D_{\mu}\psi_{\nu} )^{+} = 0
$$
The moduli space $\CM$
of solutions to this equation up to the gauge transformations
is finite-dimensional (for a given topological sector). Let
$m^{I},  I =1, \dots , {\rm dim}{\CM}$ be some choice of the
local coordinates on $\CM$. The path integral becomes an
integral over the space of loops in $\CM$. To write
an effective Lagrangian for the quantum mechanics we need
a description of the geometry of the moduli space (the
description is similar to the one presented in \HarStro,  
\bjsv).

Let us consider a particular solution $A$ (a connection
on a vector bundle $E$) of the equation $F^{+}=0$
and take its deformation $A^{\prime} = A + a$. It satisfies
the linearized equations $D_{A}^{+} a = 0$. The trivial  
solutions
to this equation $a = D_{A} \epsilon$ are the  
infinitesimal gauge
transformations. The true tangent space $T_{A}{\CM}$ is
the first cohomology group
$H^{1} = {\rm Ker} D_{A}^{+} / {\rm Im} D_{A}$
of the Atiyah-Hitchin-Singer complex:
\eqn\ahsc{\matrix{
&  & &  & D_{A}  &   & D_{A}^{+} &   &  &  &\cr
& 0 & \rightarrow &
\Omega^{0}(X, adE)
& \rightarrow &
\Omega^{1}(X, adE)  & \rightarrow &
\Omega^{2,+}(X, adE)  & \rightarrow & 0 &\cr}}
For the moduli space to be smooth one needs other cohomology
groups to vanish. More generally, the compactified moduli
space is a stratified space (stack)
whose strata have constant dimensions
of all cohomology groups of \ahsc. This allows to construct
the (formal) characteristic classes of the tangent bundle,
defined as an element of the $K$-functor of $\CM$.
Now assume that the virtual dimension of ${\CM}$ (which is the
index of the complex \ahsc) coincides with the dimension of
$H^{1}$. We can choose a basis in this space and represent
it by the solutions to the linearized instanton equations,
which obey some gauge fixing condition. A convenient choice of
gauge is $D_{A}^{*} a = 0$. So, let us pick a set of
linearly independent solutions to the equations:
\eqn\lnrasd{ D_{A}^{+} a_{I} = 0 \qquad D_{A}^{*} a_{I} = 0 }
for $I = 1 \dots , {\rm dim} {\CM}$.
The metric on the moduli space has the form:
\eqn\mtrc{G_{IJ} = \int_{X} a_{I} \wedge * a_{J}}

Now let us consider a gauge field, whose space-like
part  $A_{\mu} (x ; t) dx^{\mu}$ solves the instanton  
equations.
There exists (locally in $\CM$) a section (a universal
instanton) ${\cal A}(m) = {\cal A}_{\mu}dx^{\mu} ( x;  
m)$. We have
\eqn\dfncn{\eqalign{
\p_{t} A_{\mu} (x; t) = & {{\p m^{I}}\over{\p t}}
{{\p {\CA}_{\mu}}\over{\p m^{I}}} \cr
{{\p {\CA}}\over{\p m^{I}}} = & a_{I} + D_{A}  
\epsilon_{I} \cr}}
where $\epsilon_{I}$ is the compensating gauge transformation
($\CA + \epsilon$
define a connection in the universal bundle $\CE$
over $\CM \times X$).

Now let us have a look at the remaining part of the action:
$$
\{ Q , R_{k} \} = F_{\mu t}^{2} + (D_{\mu} \varphi)^{2} +
\psi^{\mu} D_{\mu} \eta + \psi^{\mu} [ D_{t} - i \varphi,  
\psi_{\mu}
]
$$
The fermion $\psi = \psi_{\mu}dx^{\mu}$ already satisfies
$D_{A}^{+}\psi = 0$ (after integrating out $\chi$) and  
the integral
over $\eta$ enforces the condition $D_{A}^{*}\psi = 0$.
Thus, $\psi = \chi^{I} a_{I}$, where $\chi^{I}$ are the  
fermionic
coordinates, tangent to $\CM$. Introduce the covariant
derivative
$$
\nabla_{I} = {{\p}\over{\p m^{I}}} + \epsilon_{I}
$$
We have:
$$
[ \nabla_{I}, D_{A} ] = a_{I}
$$
Clearly, $d_{\CM} \sim \chi^{I} \nabla_{I}$. We also  
have: $F_{\mu t}
= a_{I} {\dot m}^{I} + D_{\mu} ( \epsilon_{I} {\dot  
m}^{I} - A_{t}
)$.
Introduce the `improved' connection
$$
B = A_{t} - \epsilon_{I} {\dot m}^{I} + i \varphi
$$

Substituting the expressions for all fields in the action  
we get:
\eqn\actniii{
S = \int_{S^{1}} G_{IJ} [ {\dot m}^{I} {\dot m}^{J} +
\chi^{I} \nabla_{t} \chi^{J} ] + \int_{X} d^{4}x \sqrt{g}
D_{\mu} B D^{\mu} {\bar B} + \psi_{\mu} [ {\bar B},  
\psi^{\mu} ]}
After integrating out
$\bar B$ we are left with the standard action of the
$N = \half$ supersymmetric quantum mechanics with $\CM$  
as a target
space. We also have:
$$
B = {1\over{\Delta}} [\psi, * \psi] = \Phi_{IJ} \chi^{I}  
\chi^{J}
$$
It is important to notice that $\Phi_{IJ}$ has a local
form, in fact it is a curvature two-form of the connection
$\nabla_{I}$ (this discussion is parallel to the one in  
\bjsv) :
$$
\Phi_{IJ} = [ \nabla_{I}, \nabla_{J} ]
$$
as follows from the identities:
$$
a_{I} = [\nabla_{I}, D_{A} ] \Rightarrow [a_{I}, *a_{J}] =
D_{A}^{*}D_{A}
\Phi_{IJ}
$$

The main conclusion is that we have reduced the path integral
in the five-dimensional theory to the one of the  
supersymmetric
quantum mechanics on $\CM$. Using the standard localization
arguments one reduces the partition function to the integral
over $\CM$ of the $\hat A$-genus density:
$$
{\hat A} ({\CM}) = {\rm det} {{{\CR}\over{4\pi i}}\over
{{\rm sinh} {{\CR}\over{4\pi i}}}}
$$

\sssec{Strong \quad coupling \quad limit.}
In the opposite limit $\alpha_{m} \to 0$, $\alpha_{a} \to  
\infty$
(which corresponds to the low energy expansion of the  
effective
four-dimensional theory) one has the potential $(D_{t}  
\varphi)^{2}$
which breaks the gauge group $G$ down to its maximal
torus $T$ (quite similarly to the conventional four  
dimensional
story). The subtlety is that the order parameter has the
periodicity property. Indeed, one can find a gauge
$$
A_{t} + i \varphi =  \sum_{i} a_{i} H_{i}
$$
with $\p_{t}a_{i}=0$, $H_{i}$
being the generators of the Cartan subalgebra $\liet$.
This gauge leaves out the discrete gauge transformations,
which act on $a_{i}$ as follows:
$$
a_{i} \to a_{i} + {2\pi i}{{n_{i}}\over{R}}, \quad n_{i}  
\in \IZ
$$
Eventually we get the four-dimensional
effective theory, described in the section $2.1$,
where the periodicity gets restored once the one-loop
prepotential coming from the massless photon(s)
is taken into account.

\listrefs

\bye